\documentclass{emulateapj}
\providecommand{\comment}[1]{{\bf(#1)}}
\renewcommand{\comment}[1]{}
\usepackage{graphicx}
\usepackage{graphics}
\usepackage{natbib}
\usepackage{aas_macros}
\usepackage{amsmath}
\usepackage{amssymb}
\usepackage{fancyvrb}
\usepackage{epsf}

\usepackage{grffile}

\shorttitle{Halo-Model Interpretation of Galaxy Clustering Evolution}
\shortauthors{Skibba et al.}

\newcounter{appfig}
\newcounter{apptab}

\begin{document}

\title{Dark Matter Halo Models of Stellar Mass-Dependent Galaxy Clustering in PRIMUS+DEEP2 at $0.2<$\lowercase{\textit{z}}$<1.2$}

\author{
Ramin~A.~Skibba\altaffilmark{1},
Alison~L.~Coil\altaffilmark{1},
Alexander~J.~Mendez\altaffilmark{1,2},
Michael~R.~Blanton\altaffilmark{3},
Aaron~D.~Bray\altaffilmark{4},
Richard~J.~Cool\altaffilmark{5},
Daniel~J.~Eisenstein\altaffilmark{4},
Hong~Guo\altaffilmark{6}, 
Takamitsu~Miyaji\altaffilmark{7,8}, 
John~Moustakas\altaffilmark{9}, 
Guangtun~Zhu\altaffilmark{2,10}
}
\altaffiltext{1}{Department of Physics, Center for Astrophysics and Space Sciences, University of California, San Diego, 9500 Gilman Dr., La Jolla, CA 92093; raminskibba@gmail.com}
\altaffiltext{2}{Department of Physics \& Astronomy, Johns Hopkins University, 3400 N. Charles Street, Baltimore, MD 21218, USA}
\altaffiltext{3}{Center for Cosmology and Particle Physics, Department of Physics, New York University, 4 Washington Place, New York, NY 10003}
\altaffiltext{4}{Harvard-Smithsonian Center for Astrophysics, 60 Garden Street, Cambridge, MA 02138, USA}
\altaffiltext{5}{MMT Observatory, 1540 E Second Street, University of Arizona, Tucson, AZ 85721, USA}
\altaffiltext{6}{Key Laboratory for Research in Galaxies and Cosmology of Chinese Academy of Sciences, Shanghai Astronomical Observatory, Shanghai 200030, China}
\altaffiltext{7}{Instituto de Astronom\'{i}a, Universidad Nacional Aut\'{o}noma de M\'{e}xico, Ensenada, Baja California, Mexico}
\altaffiltext{8}{Visiting Scholar, University of California, San Diego}
\altaffiltext{9}{Department of Physics and Astronomy, Siena College, 515 Loudon Road, Loudonville, NY 12211, USA}
\altaffiltext{10}{Hubble Fellow}

\begin{abstract}
We utilize $\Lambda$CDM halo occupation models of galaxy clustering to investigate the evolving stellar mass dependent clustering of galaxies in the PRIsm MUlti-object Survey (PRIMUS) and DEEP2 Redshift Survey over the past eight billion years of cosmic time, between $0.2<z<1.2$. 
These clustering measurements provide new constraints on the connections between dark matter halo properties and galaxy properties in the context of the evolving large-scale structure of the universe.  
Using both an analytic model and a set of mock galaxy catalogs, we find a strong correlation between central galaxy stellar mass and dark matter halo mass over the range $M_\mathrm{halo}\sim10^{11}$-$10^{13}~h^{-1}M_\odot$, approximately consistent with previous observations and theoretical predictions.  However, the stellar-to-halo mass relation (SHMR) and the mass scale where star formation efficiency reaches a maximum appear to evolve more strongly than predicted by other models, including models based primarily on abundance-matching constraints. 
We find that the fraction of satellite galaxies in haloes of a given mass decreases significantly 
from $z\sim0.5$ to $z\sim0.9$, partly due to the fact that haloes at fixed mass are rarer at higher redshift and have lower abundances. We also find that the $M_1/M_\mathrm{min}$ ratio, a model parameter that quantifies the critical mass above which haloes host at least one satellite, decreases from $\approx20$ at $z\sim0$ to $\approx13$ at $z\sim0.9$.  
Considering the evolution of the subhalo mass function vis-\`{a}-vis satellite abundances, this trend has implications for relations between satellite galaxies and halo substructures and for intracluster mass, which we argue has grown due to stripped and disrupted satellites between $z\sim0.9$ and $z\sim0.5$. 
\end{abstract}


\section{Introduction}\label{sec:intro}

In the $\Lambda$CDM cosmology, structures form hierarchically, such that smaller haloes merge to form larger and more massive haloes.  All galaxies are thought to form as a result of gas cooling at the center of the potential well of dark matter haloes.  When a halo and its `central' galaxy are accreted by a larger halo, it becomes a subhalo and its galaxy becomes a `satellite' galaxy.
In addition to mergers, haloes also grow by smooth accretion and galaxies grow by \textit{in situ} star formation when fuel (i.e., cool gas) is available. 
In this paradigm of hierarchical structure formation, there is a correlation between halo formation and abundances and the surrounding large-scale structure such that more massive haloes tend to reside in denser regions (Mo \& White 1996).  Galaxy formation models assume that the properties of a galaxy are determined entirely by the mass and formation history of the dark matter halo within which it formed. Thus, the correlation between halo properties and environment induces a correlation between galaxy properties and environment.

The halo model is a useful framework for discussing how galaxy clustering depends on the properties of the galaxies' parent dark matter haloes, and it is a useful guide for studying the connections between galaxy formation and halo assembly (see Cooray \& Sheth 2002; Mo, van den Bosch \& White 2010 for a review). 
Halo models of galaxy abundances and clustering generally consist of the following three types: 
halo occupation distributions (HODs; e.g., Seljak 2000; Scoccimarro et al.\ 2001; Berlind \& Weinberg 2002; Kravtsov et al. 2004), conditional luminosity functions (CLFs; e.g., Yang et al.\ 2003; Cooray 2006), and (sub)halo abundance matching (SHAMs; Vale \& Ostriker 2006; Conroy, Wechsler \& Kravtsov 2006). 
Such models are useful for exploring the relations between galaxy formation and dark matter halo assembly in the context of the large-scale structure of the Universe.  The stellar to halo mass relation (SHMR) is commonly studied one in the literature (Mandelbaum et al.\ 2006; Moster et al.\ 2010; Behroozi et al.\ 2010), and one of the goals of this paper is to analyze its evolution.\footnote{Note that these are sometimes mistakenly referred to as `abundance matching relations' though they can be inferred with a variety of methods: for example, early HOD models produced luminosity-halo mass relations (e.g., Peacock \& Smith 2000), and one could argue that early constraints on galaxy-halo relations were obtained with the morphology-density relation (Dressler 1980; Postman \& Geller 1984) and galaxy kinematics (Zaritsky et al.\ 1993; Carlberg et al.\ 1996).}  The SHMR and its variants quantify the fundamental correlation between the size of central galaxies and the parent haloes that host them, spanning from low-mass dwarfs and Milky Way galaxies to galaxies in massive groups and clusters, which are hosted by massive haloes. 
The ratio has also been used to define the halo mass scale of maximum star formation efficiency in galaxies, as it provides a ratio of baryons to dark matter as a function of halo mass.

In addition to these relations, we are also interested in exploring the \textit{distributions} of galaxy and halo properties with more complex models (e.g., Skibba \& Sheth 2009; Hearin \& Watson 2013; Cohn \& White 2013). 
Moreover, sophisticated self-consistent evolutionary models using star formation histories and merger trees have been recently developed (e.g., Behroozi et al.\ 2013c; Yang et al.\ 2013), and these are complementary to semi-analytic models as well (e.g., Kang et al.\ 2012; Q.\ Guo et al.\ 2013; Contreras et al.\ 2013; Campbell et al.\ 2014). 

The literature contains an impressive array of work applying halo models to measurements of evolving galaxy clustering and lensing, and our work is complementary to them. 
For example, galaxy clustering has been modeled at a wide range of redshifts beyond $z\sim0.2$ in 
the SDSS and BOSS surveys (Wake et al.\ 2008; White et al.\ 2011; H. Guo et al.\ 2014); 
DEEP2 (Conroy et al.\ 2006; Zheng et al.\ 2007; Watson \& Conroy 2013); 
VVDS (Abbas et al.\ 2010); 
COSMOS (Leauthaud et al.\ 2012; Tinker et al.\ 2013; McCracken et al.\ 2015); 
VIPERS (de la Torre et al.\ 2013); 
CFHTLenS (Coupon et al.\ 2015); 
and Spitzer SPT (Martinez-Manso et al.\ 
2015)\footnote{Details about these surveys can be found at the following references, respectively: York et al.\ (2000); Dawson et al.\ (2013); Davis et al.\ (2003); Le F\'{e}vre et al.\ (2005); Scoville et al.\ (2007); Guzzo et al.\ (2014); Heymans et al.\ (2012); Ashby et al.\ (2013).}.  
Other authors have analyzed group catalogs and density field reconstruction as well.
However, previous work at $z\sim1$ has been often limited to relatively small samples in volumes subject to substantial`cosmic variance' errors 
and/or lacked accurate stellar masses or spectroscopic redshifts. 

In this paper, we utilize data from the PRIsm MUlti-object Survey (PRIMUS; Coil et al.\ 2011; Cool et al.\ 2013), using volume-limited catalogs constructed from a parent sample of over ∼130,000 galaxies with robust redshifts in seven independent fields covering 9~${\rm deg}^2$ of sky. 
In Skibba et al.\ (2014; hereafter S14), we modeled the optical luminosity and color of galaxy clustering at $0.2<z<1.0$ with a simple halo model in which the HOD parameters are assumed to be constant with redshift. 
Using new clustering measurements as a function of stellar mass and specific star formation rate (Mendez et al., in prep.; hereafter M15), we now analyze the scale-dependent clustering evolution with improved halo models to study the evolving relations between stellar mass  and dark matter halo mass of central and satellite galaxies. 
In order to perform a complete analysis and to obtain robust model parameters, we analyze the spatial distributions and abundances of PRIMUS galaxies with two types of models: an analytic model and a set of mock galaxy catalogs. 
The former is based on dark matter halo statistics and quantities measured from numerical simulations, including the mass function, bias function, and density profile. The latter is directly tied to a dark matter cosmological simulation and halo-finding algorithm. These complementary methods are both widely used in the literature.

This paper is organized as follows. 
In Section~\ref{sec:PRIMUSCFs}, we briefly describe the PRIMUS galaxy clustering measurements of M15.  
We describe the analytic halo model of galaxy clustering and mock galaxy catalogs in Sec.~\ref{sec:model}, and additional details are described in the appendices. 
We then present our HOD model results in Sec.~\ref{sec:results}, and we discuss the results and their implications, such as for the stellar mass-halo mass relation and satellite abundances, in Sec.~\ref{sec:implications}. 
Finally, we end with a summary of our conclusions in Sec.~\ref{sec:conc}.

Throughout this paper, we adopt a flat $\Lambda$CDM cosmology with
$\Omega_m=0.27$, $\Omega_\Lambda=1-\Omega_m$, 
$\sigma_8=0.80$\footnote{Note that these values of $\Omega_m$ and $\sigma_8$ are slightly lower than the latest cosmological constraints (Planck collaboration et al.\ 2014).}, $n_s=1$, and we express units that depend on the Hubble
constant in terms of $h \equiv H_0/100~{\rm km}\,{\rm s}^{-1}\,{\rm Mpc}^{-1}$.  
For the halo masses and radii, we define them with a spherical overdensity 200 times the mean density of the Universe (see Appendix~\ref{app:Mdef}), except when specified otherwise.  
$M$ or $M_h$ refers to halo mass and $m$ refers to subhalo mass (in units of $h^{-1}M_\odot$), and $M_\ast$ refers to stellar mass ($M_\odot$ units).
For the stellar masses and star formation rates (SFRs), we assume 
a universal Chabrier (2003) initial mass function (IMF). 

\section{Galaxy Clustering Measurements}\label{sec:PRIMUSCFs}

In S14, we presented spatial clustering measurements of galaxies with high-quality redshifts in the PRIMUS and DEEP2 surveys as a function of luminosity and color over the redshift range $0.2<z<1.2$. 
In M15, we present clustering measurements as a function of stellar mass and specific star formation rate (sSFR) over the same redshift range, and these are the measurements we use here. 
The clustering measurements build on the stellar mass-dependent clustering of active galactic nuclei (AGN) in Mendez et al.\ (2015a). 
We quantify galaxy clustering with two-point projected auto-correlation functions, $w_p(r_p)$, which are measured by integrating $\xi(r_p,\pi)$ out to line-of-sight separations of $\pi_\mathrm{max}=80~h^{-1}\mathrm{Mpc}$ for PRIMUS fields and $20~h^{-1}\mathrm{Mpc}$ for DEEP2. 
The catalogs are roughly volume-limited, and incompleteness and redshift success weights are applied similarly as in S14. The correlation function errors are estimated with jackknife resampling methods, which are described in S14 and M15.

\begin{figure}
   	\includegraphics[width=1.0\linewidth]{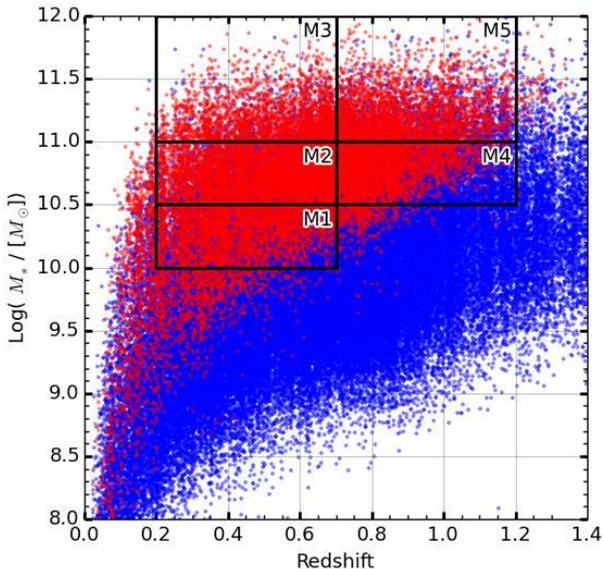} 
 		\caption{Stellar mass and redshift limits of the five PRIMUS samples described in Table~\ref{tab:samples}. Blue and red points indicate star-forming and quiescent galaxies based on the SFR-$M_\ast$ demarcation of Moustakas et al.\ (2013).}
 	\label{fig:Mzplot}
\end{figure}

The main properties of the stellar mass-limited samples are listed in Table~\ref{tab:samples}, and the stellar mass and redshift limits and distributions are shown in Figure~\ref{fig:Mzplot}.  
M15 includes the PRIMUS science fields (CDFS-SWIRE, XMM-LSS, COSMOS, DLS, and ES1) as well as the Extended Groth Strip (Lin et al.\ 2007) and DEEP2 fields. 
We refer the reader to M15 for details.

\begin{table*}
\caption{Properties of the Stellar-Mass Limited Samples}
 \centering
 \begin{tabular}{ l | c c c c c c c c }
   \hline
   sample & $z_\mathrm{min}$ & $\langle z\rangle$ & $z_\mathrm{max}$ & $\mathrm{log}M_{\ast,\mathrm{min}}$ & $\langle\mathrm{log}M_\ast\rangle$ & $\mathrm{log}M_{\ast,\mathrm{max}}$ & $\mathrm{log}~\bar n_\mathrm{gal}$ & $b_\mathrm{gal}$ \\
   \hline 
   M1 & 0.20 & 0.49 & 0.70 & 10.00 & 10.65 & 12.00 & -2.37$\pm$0.06 & 1.35$\pm$0.02 \\ 
   M2 & 0.20 & 0.50 & 0.70 & 10.50 & 10.87 & 12.00 & -2.69$\pm$0.07 & 1.52$\pm$0.06 \\ 
   M3 & 0.20 & 0.51 & 0.70 & 11.00 & 11.20 & 12.00 & -3.45$\pm$0.11 & 1.74$\pm$0.15 \\ 
   M4 & 0.70 & 0.87 & 1.20 & 10.50 & 10.90 & 12.00 & -2.7$\pm$0.1   & 2.13$\pm$0.16 \\ 
   M5 & 0.70 & 0.91 & 1.20 & 11.00 & 11.20 & 12.00 & -3.35$\pm$0.07 & 2.5$\pm$0.3 \\ 
   \hline
  \end{tabular}
  \begin{list}{}{}
    \setlength{\itemsep}{0pt}
    \item Stellar mass threshold samples: redshift limits, stellar mass limits ($M_\odot$), and number densities (in units of $10^{-3}h^3\mathrm{Mpc}^{-3}$), and bias. 
 Mean redshifts and stellar masses of the samples are listed, and their median values are not significantly different. The lower error bars of the high-redshift biases indicate the results without the COSMOS field (see Appendix~\ref{app:withoutcosmos} and M15 for details). 
All of the values quoted are obtained from M15, except for the number densities and their errors, which are obtained from Moustakas et al.\ (2013; Table~5) and are used here as additional constraints. 
  \end{list}
  \label{tab:samples}
\end{table*}

The stellar masses and SFRs are taken from Moustakas et al.\ (2013), and they are estimated from multi-wavelength imaging from UV to mid-IR wavelengths using a Bayesian spectral energy distribution (SED) modeling code (\texttt{iSEDfit}). 
The fiducial parameters are based on the flexible stellar population synthesis models of Conroy \& Gunn (2010), and exponentially declining star formation histories (SFHs) with Gaussian bursts of star formation superposed are assumed. 
Metallicities near the solar value (Asplund et al.\ 2009) are assumed, and the time-dependent dust attenuation curve of Charlot \& Fall (2000) is adopted.  The number densities and errors in Table~\ref{tab:samples} are computed from the stellar mass functions (SMFs) of Moustakas et al.\ (2013), which includes a thorough analysis of relevant systematic uncertainties.

\section{Model Description}\label{sec:model}

In this section, we describe the components of our halo model of galaxy clustering.  
We will describe the halo occupation distribution (HOD) model, the galaxy clustering model, and the method for constructing mock catalogs. 
Except when stated otherwise, we utilize the same model components in both the analytic model and the model for constructing mock catalogs.

Throughout this section, $M$ refers to the mass of dark matter haloes\footnote{Note that the same formalism may be utilized to model galaxy clustering as a function of halo circular velocity (e.g., $V_\mathrm{max}$), rather than mass.} (defined using a virial overdensity 200 times the mean density of the Universe, unless stated otherwise) and $M_\ast$ refers to the stellar mass of galaxies. 

\subsection{Halo Occupation Distribution}\label{sec:HOD}

For these calculations, we use the mean redshifts of the observed galaxy samples and their number densities in addition to the projected clustering measurements. 
For the analytic model, we assume a Tinker et al.\ (2008b) halo mass function, Tinker et al.\ (2010) halo bias, 
and Eisenstein \& Hu (1998) matter power spectrum.  
For both the analytic model and mock galaxy catalogs, we assume a Navarro, Frenk \& White (1997) density profile. 
The details of these quantities are described in Appendix~\ref{app:modelstuff}.

We use a model of the halo occupation distribution (HOD) similar to that in Skibba \& Sheth (2009) and Zheng et al.\ (2007), and we refer the reader to these papers for details. 
The mean \textit{central} galaxy HOD is modeled with the following parametrization:
\begin{equation}
  \langle N_\mathrm{cen}|M,M_\ast\rangle \,=\, \frac{1}{2}\Biggl[1\,+\,\mathrm{erf}\Biggl(\frac{\mathrm{log}(M/M_\mathrm{min})}{\sigma_{\mathrm{log}M}}\Biggr)\Biggr]
 \label{eq:NcenM}
\end{equation}
where $\mathrm{erf}$ is the error function, which assumes a lognormal distribution for the central galaxy conditional SMF,  
\begin{eqnarray}
\Phi_{\rm cen}(M_\ast|M) d M_\ast &=& {1 \over \sqrt{2\pi}\, {\rm ln}(10) \, \sigma_{{\rm log}M}} \\
&& \exp\left[ -\left({\log(M_\ast/M_{\ast,\rm cen})\over\sqrt{2}\sigma_{{\rm log}M}}\right)^2\right] \, {d M_\ast \over M_\ast}\,. \nonumber 
 \label{eq:PcenM}
\end{eqnarray}
The mean \textit{satellite} galaxy HOD is parametrized by 
\begin{equation}
  \langle N_\mathrm{sat}|M,M_\ast\rangle = 
    \Biggl(\frac{M-M_0}{M_1^{ ' }}\Biggr)^\alpha ,
 \label{eq:NsatM}
\end{equation}
and we discuss the satellite conditional SMF in Section~\ref{sec:sats}.

Note that unlike the simple model in S14, in this paper we allow the HOD parameters to evolve with redshift. 
The halo mass (threshold) and stellar mass (threshold) are directly related as the $M_\ast$-$M_\mathrm{halo}$ relation, $\sigma_{\mathrm{log}M}$ determines the lognormal scatter in the relation, the $\mu=M_1/M_\mathrm{min}$ parameter determines the critical mass above which haloes typically host at least one satellite within the selection limits, and $\alpha$ is the power-law index of the mass dependence of the efficiency of satellite galaxy formation. 
$M_0$ describes the smooth drop-off of the satellite HOD at low halo mass (relative to the mass threshold), but in practice a model with $(M/M_1)^\alpha$ produces similar results (Zheng et al.\ 2007)\footnote{$(M/M_1)^\alpha$ is the way the mean satellite HOD has been traditionally modeled, so we write $M_1^{ ' }$ in (\ref{eq:NsatM}) because of the $M_0$ parameter.}.

$\langle N|M\rangle$ is the mean HOD of \textit{all} galaxies (i.e., centrals and satellites),  
\begin{equation}
 \langle N_\mathrm{gal}|M,M_\ast\rangle \,\equiv\, \langle N_\mathrm{cen}|M,M_\ast\rangle ( 1 + \langle N_\mathrm{sat}|M,M_\ast\rangle )
 \label{eq:NgalM}
\end{equation}
The mean galaxy number density, $\bar n_{\rm gal}$, is defined in the analytic model with the following: 
\begin{eqnarray}
  \bar n_{\mathrm {gal}}(M_\ast) &=& \int_{M_{\rm min}(M_\ast)}^{M_{\rm max}(M_\ast)} dM \, {dn(M,z)\over dM} \nonumber\\ 
    && \ \times \langle N_{\mathrm {cen}}|M\rangle\,\Bigl[1 + \langle N_{\mathrm {sat}}|M\rangle\Bigr]
 \label{eq:ngal}
\end{eqnarray}

For the central galaxies, we are implicitly assuming that their stellar masses have a lognormal distribution at fixed halo mass, though it is not known how accurate this assumption is (More, Diemer \& Kravtsov 2015).  
For the satellite HOD, we are assuming that they follow a Poisson distribution (Kravtsov et al.\ 2004; Zheng et al.\ 2005).  We perform most of our model analysis without allowing $M_0$ to be a free parameter, as it is poorly constrained and we are interested in exploring constraints on more important satellite HOD parameters such as $M_1=\mu M_\mathrm{min}$. 

\begin{figure}[h!]
 \includegraphics[width=1.0\linewidth]{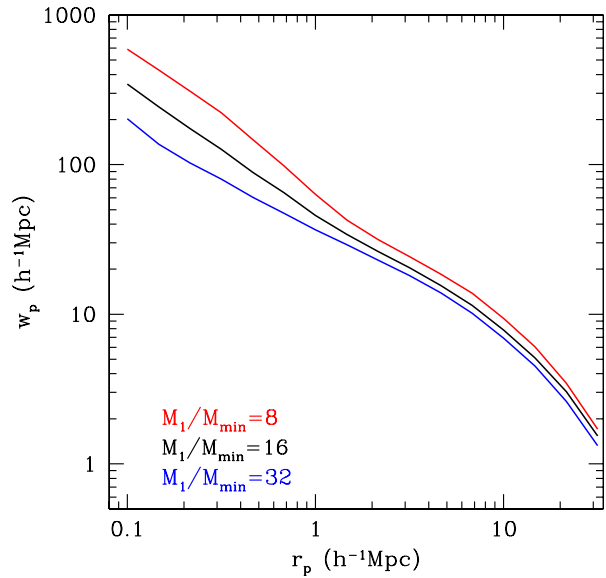} 
 \caption{An example of the dependence of model correlation functions on the $M_1/M_\mathrm{min}$ parameter for a given $M_\mathrm{min}$ (i.e., assuming the same $M_\ast$-$M$ relation).  
 Clustering predictions of mock galaxy catalogs with $M_1/M_\mathrm{min}=8$, 16, and 32 are shown, spanning the range of allowed parameter values.  Lower values of this parameter imply a higher satellite fraction and more central-satellite and satellite-satellite pairs. 
 }
 \label{fig:M1Mmin}
\end{figure}

The example correlation functions in Figure~\ref{fig:M1Mmin} show how the clustering varies with $M_1/M_\mathrm{min}$ (Eq.~\ref{eq:NsatM}) over the maximum range of the parameter. 
Note that lower values of the parameter imply more satellites and larger number densities, but $\bar n_{\rm gal}$ is primarily determined by the halo mass threshold. 
There is some degeneracy between $M_1/M_\mathrm{min}$ and $\alpha$, because the larger values of the latter also increases the clustering strength, though in a different scale-dependent manner. In general, when there are more satellites hosted by halos of some mass, there will be more central-satellite and satellite-satellite pairs and therefore stronger two-point clustering. 

\subsection{Galaxy Clustering Model}

In this section, we describe in more detail how galaxy clustering is modeled in the analytic model.

Galaxies and haloes are biased tracers of the underlying distribution of dark matter (e.g., Weinberg et al.\ 2004; Zehavi et al.\ 2011; S14). 
In the $\Lambda$CDM theory of hierarchical structure formation, the large-scale clustering of halos with respect to matter can be described with the \textit{halo} bias parameter: 
  $\xi_{hh}(r,M,z) \approx [b_{\rm halo}(M,z)]^2\xi_{mm}(r,z)$
(Mo \& White 1996; Sheth \& Lemson 1999), where the matter correlation function is obtained from the linear or nonlinear power spectrum (Efstathiou, Bond \& White 1992; Smith et al.\ 2003).  
In the halo model of galaxy clustering, \textit{galaxy} bias $b_{\rm gal}$ can then be inferred from the abundance and bias of halos, combined with the occupation distribution of galaxies in the halos: 
\begin{equation}
  b_{\rm gal}(M_\ast) = \int_{M_{\rm min}(M_\ast)}^{M_{\rm max}(M_\ast)} dM \frac{dn(M,z)}{dM} b_{\rm halo}(M,z) \frac{\langle N_{\rm gal}|M\rangle}{\bar n_{\rm gal}}
 \label{eq:HMbgal}
\end{equation}
(Cooray \& Sheth 2002; Yang et al.\ 2003), where the integration limits are related to the (e.g. stellar mass dependent) selection of the galaxies themselves.  
In terms of the large-scale galaxy correlation function, which depends on galaxy stellar mass $M_\ast$, 
galaxy bias can be described as 
\begin{equation}
 \xi_{gg}(r|M_\ast,z) = [b_{\rm gal}(M_\ast,z)]^2 \xi_{mm}(r,z) .
 \label{eq:bgal}
\end{equation}

In this paper, we perform a more detailed analysis using the full spatial correlation function of galaxies at projected separations of $0.1<r_p<30~h^{-1}\mathrm{Mpc}$ (i.e., we are not limiting the analysis to large scales as was done in S14). 
Following Skibba \& Sheth (2009), we perform our halo model calculations in Fourier space.  
The two-point correlation function is the Fourier transform of the 
power spectrum
\begin{equation}
  \xi(r) = \int {dk\over k}\, {k^3 P(k)\over 2\pi^2}\, {\sin kr\over kr}.
 \label{xiFT}
\end{equation}
In the halo model, $P(k)$ is written as the sum of two terms:
one that arises from galaxies within the same halo and dominates on 
small scales (the 1-halo term), and the other from galaxies in 
different halos which dominates on larger scales (the 2-halo term).  
That is,
\begin{equation}
  P(k) = P_{1h}(k) + P_{2h}(k),
 \label{1h2h}
\end{equation}
where,
\begin{eqnarray}
  P_{1h}(k|M_\ast) &=& \int_{M_\mathrm{min}(M_\ast)}^{M_\mathrm{max}(M_\ast)} dM \,\frac{dn(M)}{dM}\,
                \langle N_{\mathrm {cen}}|M\rangle \nonumber \\
	&& \ \times \Biggl[ \frac{2\,\langle N_{\mathrm {sat}}|M\rangle\,
	   u_{\mathrm {gal}}(k|M)}{\bar n_{\mathrm {gal}}^2} \nonumber \\
       &\phantom{=}& \qquad +\, 
           \frac{{\langle N_{\mathrm {sat}}(N_{\mathrm {sat}}-1)|M\rangle}\,
	  u_{\mathrm {gal}}(k|M)^2}{\bar n_{\mathrm {gal}}^2}  \Biggr], 
  \label{eq:Pk1h} \\
  P_{2h}(k|M_\ast) &=& \Biggl[ \int_{M_\mathrm{min}(M_\ast)}^{M_\mathrm{max}(M_\ast)} dM \,\frac{dn(M)}{dM}\,
                 \langle N_{\mathrm {cen}}|M\rangle
  \label{eq:Pk2h} \\
       &\phantom{=}& \;\; \times \, \frac{1\,+\,
          \langle N_{\mathrm {sat}}|M\rangle\,
	  u_{\mathrm {gal}}(k|M)}{\bar n_{\mathrm {gal}}}\,b(M) \Biggr]^2 \,
	  P_{\mathrm {lin}}(k), \nonumber
\end{eqnarray}
and $u_{\mathrm {gal}}(k|M)$ is the Fourier transform of the galaxy 
density profile, which is closely related to the subhalo and dark matter density profile. 
The occupation distribution $p_{\rm sat}(N_{\rm sat})$ is  
well-approximated by a Poisson distribution (e.g., Kravtsov et al. 2004; 
Yang et al. 2008; Wetzel \& White 2010), 
\begin{equation}
 P(N_\mathrm{sat}|M) = \frac{\lambda^{N_\mathrm{sat}}\mathrm{e}^{-\lambda}}{N_\mathrm{sat}!}
\end{equation}
where $\lambda = \langle N_\mathrm{sat}|M\rangle$, 
so we set 
$\langle N_{\mathrm {sat}}(N_{\mathrm {sat}}-1)|M\rangle\,
 =\,{\langle N_{\mathrm {sat}}|M\rangle}^2$.
The two parts of the 1-halo term in equation~(\ref{eq:Pk1h}) can be 
thought of as the `center-satellite term' and the `satellite-satellite term'.

See Appendix~\ref{app:modelstuff} for more details about the modeling of the galaxy density profile, halo bias, halo mass function, and matter power spectrum.  

\subsection{Mock Galaxy Catalogs}\label{sec:mocks}

In this section, we describe how we model `mock' galaxy catalogs, which is distinct from our analytic model and provides an alternative methodology to constrain central and satellite galaxy properties and their relations with halo mass and halocentric position.  The results of both models will be presented in Section~\ref{sec:results}. 

We construct mock galaxy catalogs using the Bolshoi dark matter simulation (Klypin et al.\ 2011)\footnote{Note that Bolshoi has $\sigma_8=0.82$ while the analytic model has $\sigma_8=0.80$.} with Rockstar phase-space halo-finding algorithm (Behroozi et al.\ 2013a, 2013b)\footnote{Note that for consistency with Tinker et al.\ (2008), we have used a $M_{200m}$ mass definition for the analytic model, while the Rockstar haloes have a $M_\mathrm{BN98}$ mass definition based on the evolving virial overdensity calibrated by Bryan \& Norman (1998).}.  
The simulation uses a computational box with length $L_\mathrm{box}=250~h^{-1}\mathrm{Mpc}$ and with impressive mass resolution (particle mass $m=1.35\times 10^8~h^{-1}M_\odot$) and force resolution ($1~h^{-1}\mathrm{kpc}$ physical). 
The model for constructing the mocks is described in Skibba et al.\ (2006) and Skibba \& Sheth (2009) with additional updates and improvements in the treatment of galaxy color distributions and dynamics that will be described in Skibba (in prep.). 
The analytic HOD model (\ref{eq:Pk1h}-\ref{eq:Pk2h}) assumes the `central galaxy paradigm', in which the most massive galaxy is assumed to be the central galaxy of a halo and at rest at the halo center, while the mock catalogs relax these assumption (see Skibba et al.\ 2011), though in practice it has only a minor effect on the small-scale projected clustering. 
Satellite galaxy distributions are assumed to follow the same (NFW) density profile as in the analytic model, and except for a test below, we ignore the subhaloes in the simulation. 
A version of this model was used to construct mock catalogs for Old et al.\ (2015). 



We will assess in Section~\ref{sec:results} how consistent the inferred HOD parameters as a function of stellar mass from the mocks are with the analytic model.  
In subsequent work, we will utilize these models to analyze the clustering and distribution of star-forming and quiescent galaxies in more detail. 

\begin{figure}
   	\includegraphics[width=1.0\linewidth]{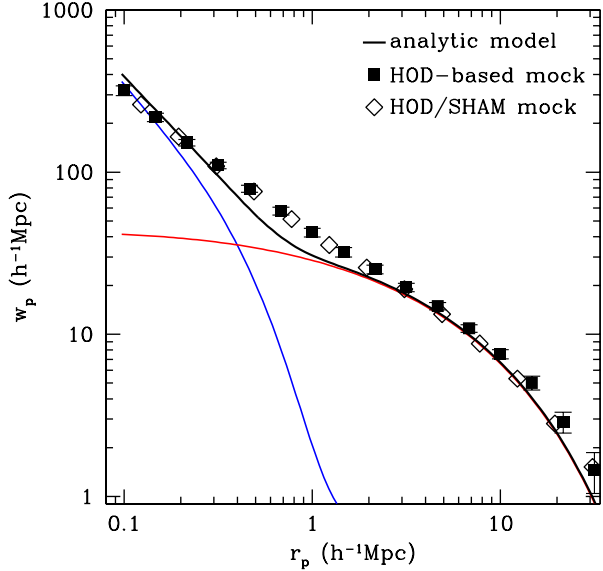} 
 		\caption{Galaxy auto-correlation function comparison for galaxies with $\mathrm{log}(M/M_\odot h^{-1})>11.4$ and $z\sim0.7$: 
                Bolshoi/Rockstar HOD-based mock galaxy catalog (square points), HOD/SHAM hybrid mock catalog (diamonds), and HOD-based analytic model (black line), including the 1-halo and 2-halo terms (blue and red lines). 
        }
 	\label{fig:modelmockwp}
\end{figure}

As an example of our models, we compare a correlation function prediction of the HOD-based analytic model and mock galaxy catalog in Figure~\ref{fig:modelmockwp}. 
The two models are sufficiently consistent where the 1-halo term dominates ($r_p<500~h^{-1}\mathrm{kpc}$) and where the 2-halo term dominates ($r_p>3~h^{-1}\mathrm{Mpc}$). 
However, the analytic model's clustering prediction for the scales in between is too low. Observed clustering measurements do exhibit a bump in this region (Zehavi et al.\ 2004), though with a more power-law-like behavior than in this analytic model (see also Watson et al.\ 2011). 
Incorporating scale dependent bias and halo exclusion, which reduce the discrepancy on these scales, are work in progress (see Appendix~\ref{app:modelstuff}), although there currently is no ideal way to treat these effects or to address the issue of halo truncation at a particular radius.

For comparison and as a test, we also show the clustering prediction of a mock catalog constructed with a HOD/SHAM hybrid model like one we developed for the comparison project in Knebe et al.\ (2015)\footnote{In practice, it is similar to a full SHAM model in which one assumes a stellar-to-halo mass relation and associates central and satellite galaxies with haloes and subhaloes in the simulation accordingly.}. In this model, satellite galaxies are directly associated with rank-ordered subhaloes and are given their positions and velocities, rather than assuming NFW distributions.  The result is nearly identical to that of our fiducial HOD-based mock catalog.  
At lower stellar masses than we probe in this paper, in the regime where a significant fraction of satellites are `orphans' whose subhaloes have been stripped away (e.g., Wang et al.\ 2006), galaxies in these two types of models have slightly different spatial distributions, however (Pujol, Skibba, \& Gazta\~{n}aga et al., in prep.).

In the next section, we will present results for both sets of models, though the constraints on the satellite HOD parameters may be more robust in the mocks because of their more realistic treatment of the 1-halo to 2-halo transition region.

\section{Results: Halo Occupation Parameters}\label{sec:results}

\subsection{Parameter-Fitting Procedure}

For both sets of models, 
we perform parameter scans and calculate $\chi^2$ values, and our procedure yields best-fit halo-model parameters and approximate 1-$\sigma$ confidence intervals. 
This is not the same as a Monte Carlo Markov Chain (MCMC) procedure (see e.g., Tinker et al.\ 2013), though such an analysis should yield nearly identical results; 
our procedure is also similar to that of Wake et al.\ (2011). 

We quantify the total $\chi^2$ for a given model and dataset with the following:
\begin{equation}
 \chi_\mathrm{tot}^2 = \sum \biggl[ \chi_\mathrm{SMF}^2 + \sum_{i=1}^{N_w} \chi_{w,i}^2 \biggr]
\label{eq:chi2}
\end{equation}
where $\chi_\mathrm{SMF}^2$ is for the number density compared to the Moustakas et al.\ (2013) SMF 
and $\chi_{w,i}^2$ is for the $w_p(r_p)$ correlation functions compared to M15 at a range of projected separations ($0.1<r_p<30~h^{-1}\mathrm{Mpc}$).
In addition to these auto-correlation functions, one could also include rank-ordered stellar mass mark correlation functions (see Skibba \& Sheth 2009; Skibba et al.\ 2013) when determining the best-fit models, but such an analysis is beyond the scope of this paper.

In most cases, the host halo masses, $\mu\equiv M_1/M_\mathrm{min}$, and $\alpha$ are the most robust parameters, while $\sigma_{\mathrm{log}M}$ and $M_0$ are less strongly constrained.  (The scatter $\sigma_{\mathrm{log}M}$ is better constrained by other statistics such as satellite kinematics and CLFs.) 

The inferred halo-model parameters can change when observed abundances (number densities) are incorporated in the calculation. 
For the bulk of this paper, we present results in which the COSMOS field is included.  
As shown in S14, because of apparently anomalous structures in this field, it yields clustering strengths and bias values much larger than other fields.  
We split our samples at the median redshift of $z=0.7$ (see Sec.~\ref{sec:PRIMUSCFs}), which partially splits a large structure in COSMOS (Scoville et al.\ 2013) and reduces this effect. 
For comparison, in Appendix~\ref{app:withoutcosmos} we include the halo-model parameters inferred when the COSMOS field is excluded.  In general, the results are well within the 1-$\sigma$ errors of those we obtain here, with the halo masses 0.05-0.10~dex lower in the high-redshift range ($0.7<z<1.2$).

Although we will treat each stellar mass and redshift bin independently for the purposes of parameter fitting, they are related to each other.  For the better constrained parameters, fitting formulae as a function of mass and redshift can be estimated from the trends we obtain.  (See Sec.~\ref{sec:implications}; see also Appendix~A2 of Skibba \& Sheth 2009 for luminosity-dependent HOD parameters.) 
In addition, we do not include the covariance matrices of the clustering measurements in the model fitting, as they have large uncertainties; including the full error covariance matrices, when they are less noisy, versus only the diagonal elements usually has only a small effect on clustering analyses (Zehavi et al.\ 2011).  
We attempt to sufficiently finely probe the parameter space so as to estimate 1-$\sigma$ errors of the inferred model parameters.  
However, there are some weak degeneracies between parameters; for example, lower $M_1/M_\mathrm{min}$ and higher $\alpha$ both increase the satellite fraction (though with different halo mass dependencies) and increase the clustering signal of central-satellite and satellite-satellite pairs (though with different scale dependencies).  

\begin{figure}
 \includegraphics[width=1.0\linewidth]{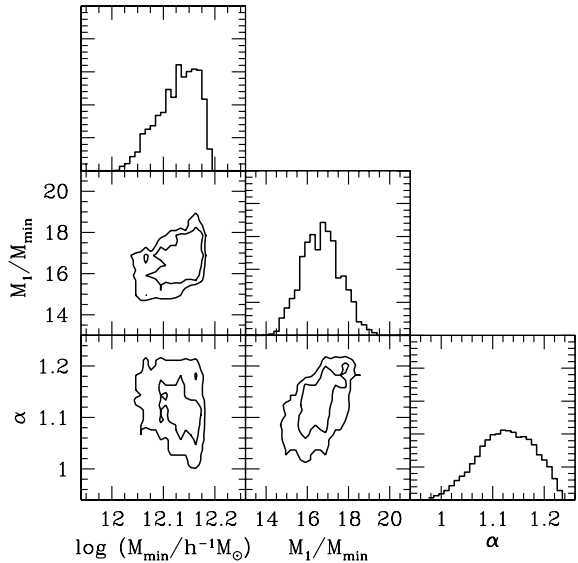} 
 \caption{1-D (diagonal) and 2-D likelihood distributions of the model parameters $M_\mathrm{min}$, $M_1/M_\mathrm{min}$, and $\alpha$ resulting from fitting the procedure for the M2 galaxies ($M_\ast\geq10^{10.5}M_\odot$ at $z\sim0.5$). 
 Distributions for the top 10\% best-fitting models are shown; the top 20\% have similar distributions. Flat priors were used within the ranges shown. 
 }
 \label{fig:paramdists}
\end{figure}

As a test, we ran models more finely through the parameter space for the M2 galaxies ($M_\ast\geq10^{10.5}M_\odot$ at $z\sim0.5$), and we show the parameter distributions we obtained for $M_\mathrm{min}$, $M_1/M_\mathrm{min}$, and $\alpha$ in Figure~\ref{fig:paramdists}). 
These demonstrate the robustness of our results, and the results shown here are very similar to those we obtain with our standard binning in the next section. 
In particular, we obtain $\mathrm{log}M_\mathrm{min}=12.132\pm0.037$, $\mu=16.80\pm0.90$, and $\alpha=1.125\pm0.053$, and in each case the means and medians are nearly identical. 
As stated above, some degeneracy between the $M_1/M_\mathrm{min}$ and $\alpha$ parameters can be observed. 
Unless stated otherwise, in the following we use halo mass bins of 0.05 dex, $M_1/M_\mathrm{min}$ bins of 1.0, $\alpha$ bins of 0.05, and $\sigma_{\mathrm{log}M}$ bins of 0.1. 

\subsection{Best-Fitting Models}


\setlength{\tabcolsep}{-2pt}
\begin{figure*}
 \centering
  \begin{tabular}{@{}ccc@{}}
    \includegraphics[width=0.35\textwidth]{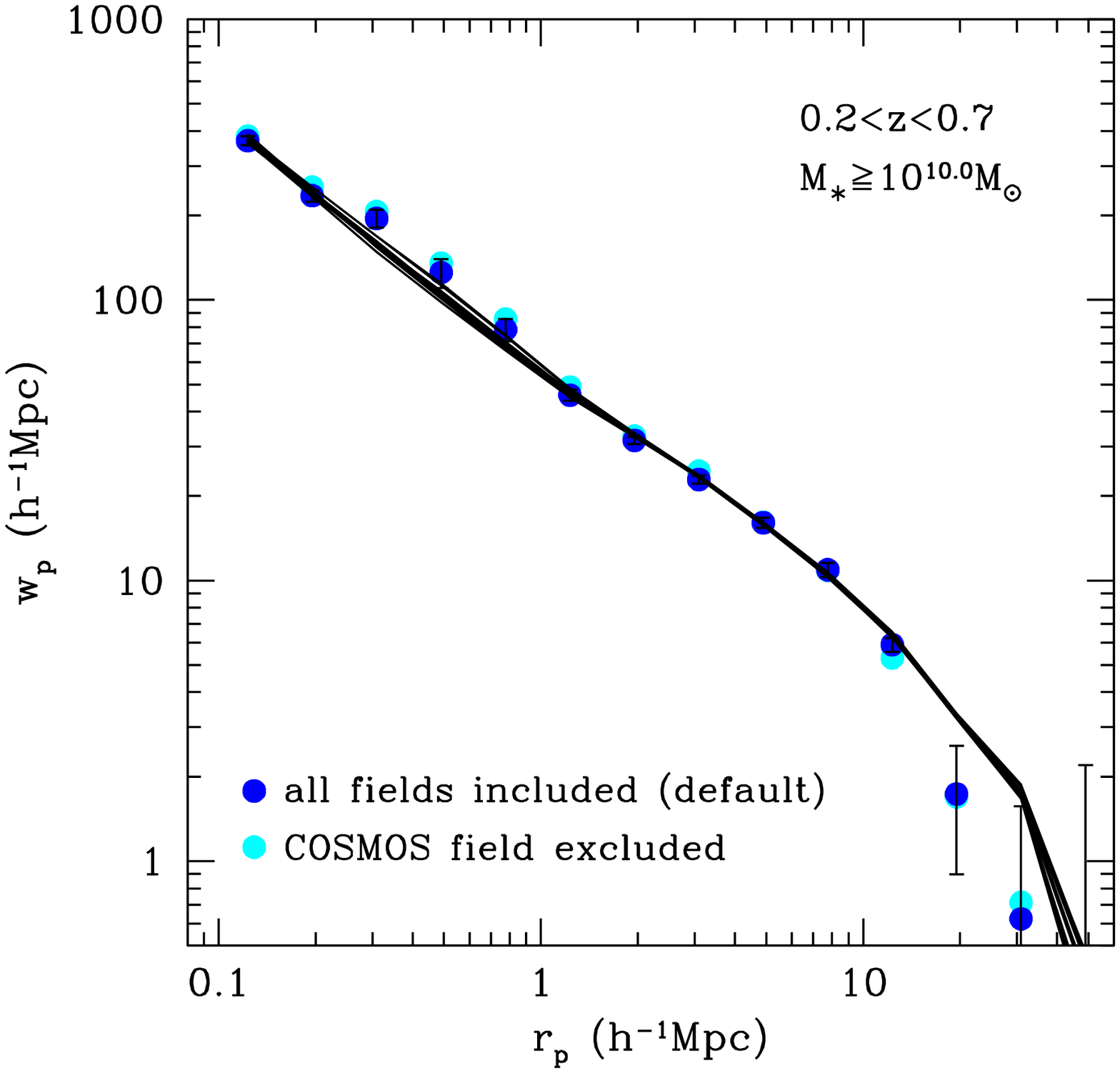} &
    \includegraphics[width=0.35\textwidth]{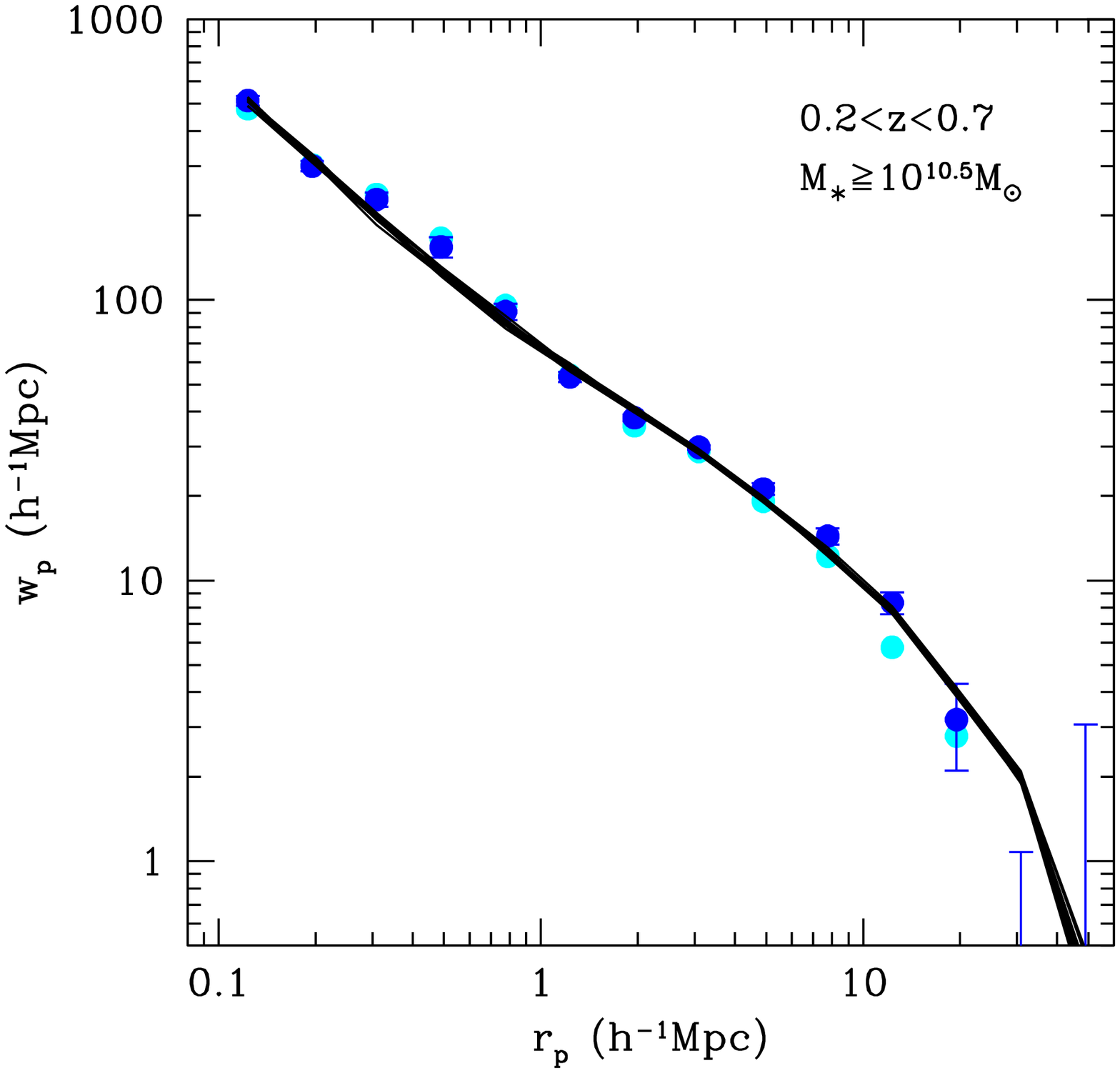} &
    \includegraphics[width=0.35\textwidth]{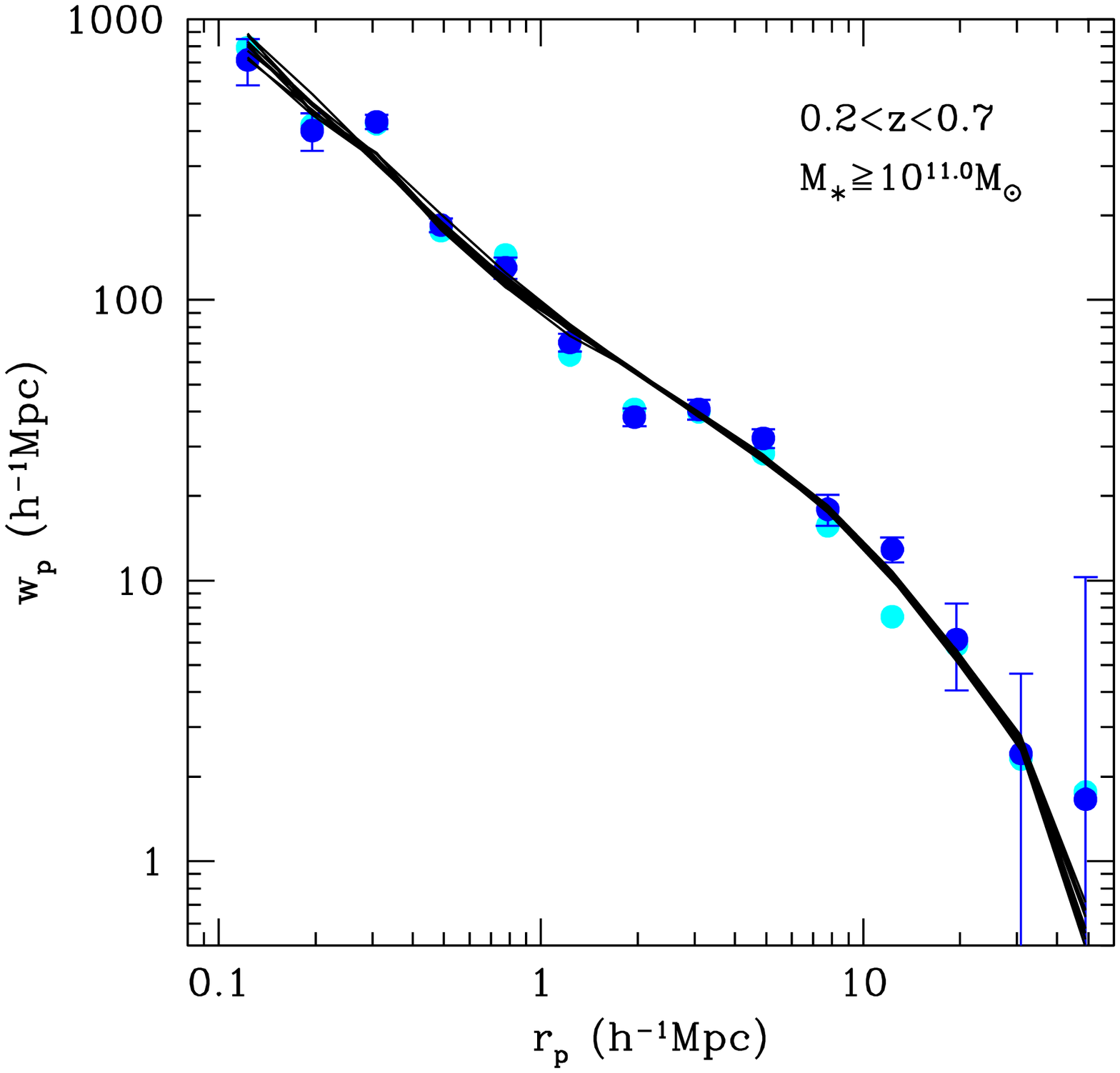} \\
    & 
    \includegraphics[width=0.35\textwidth]{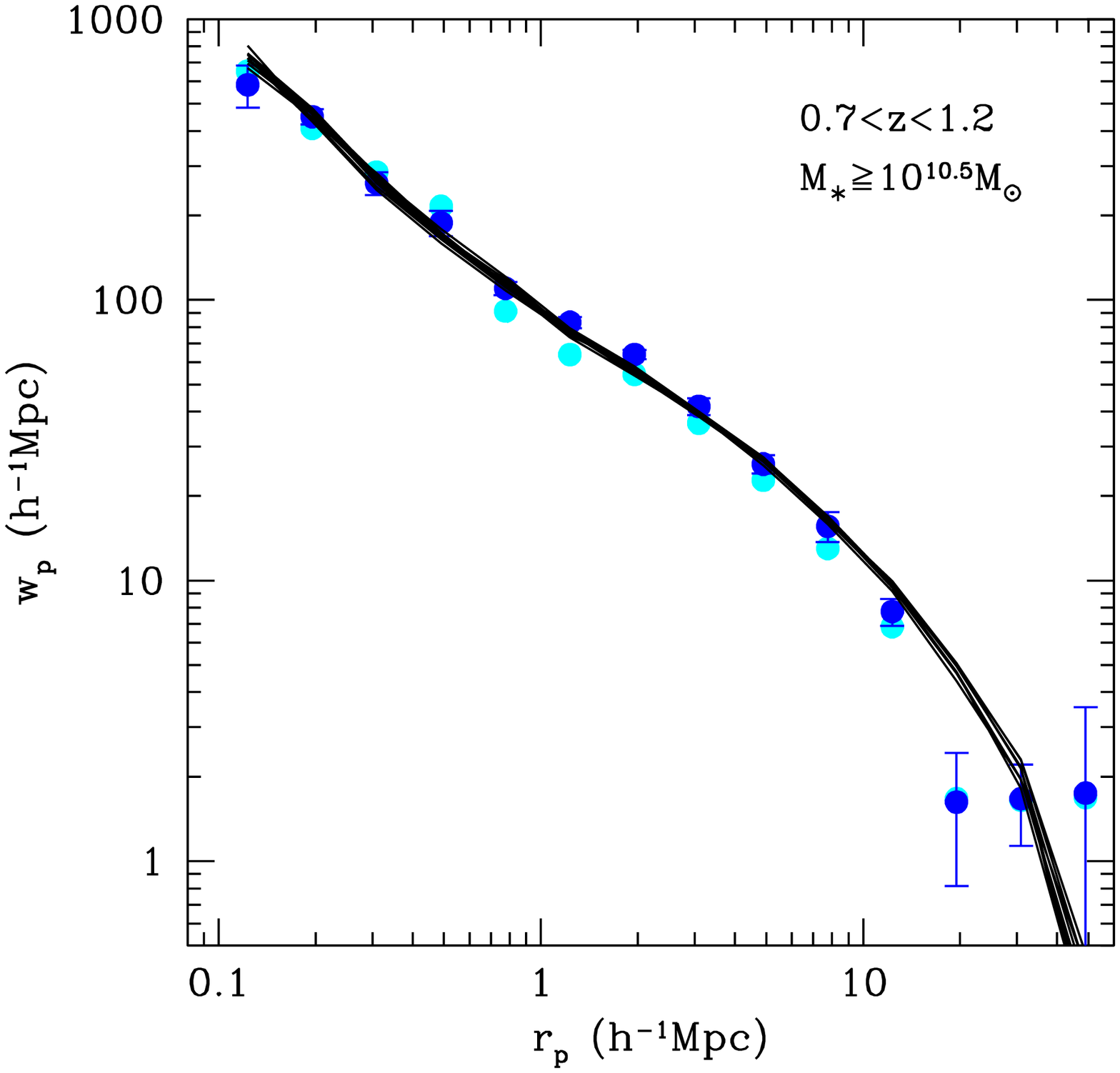} &
    \includegraphics[width=0.35\textwidth]{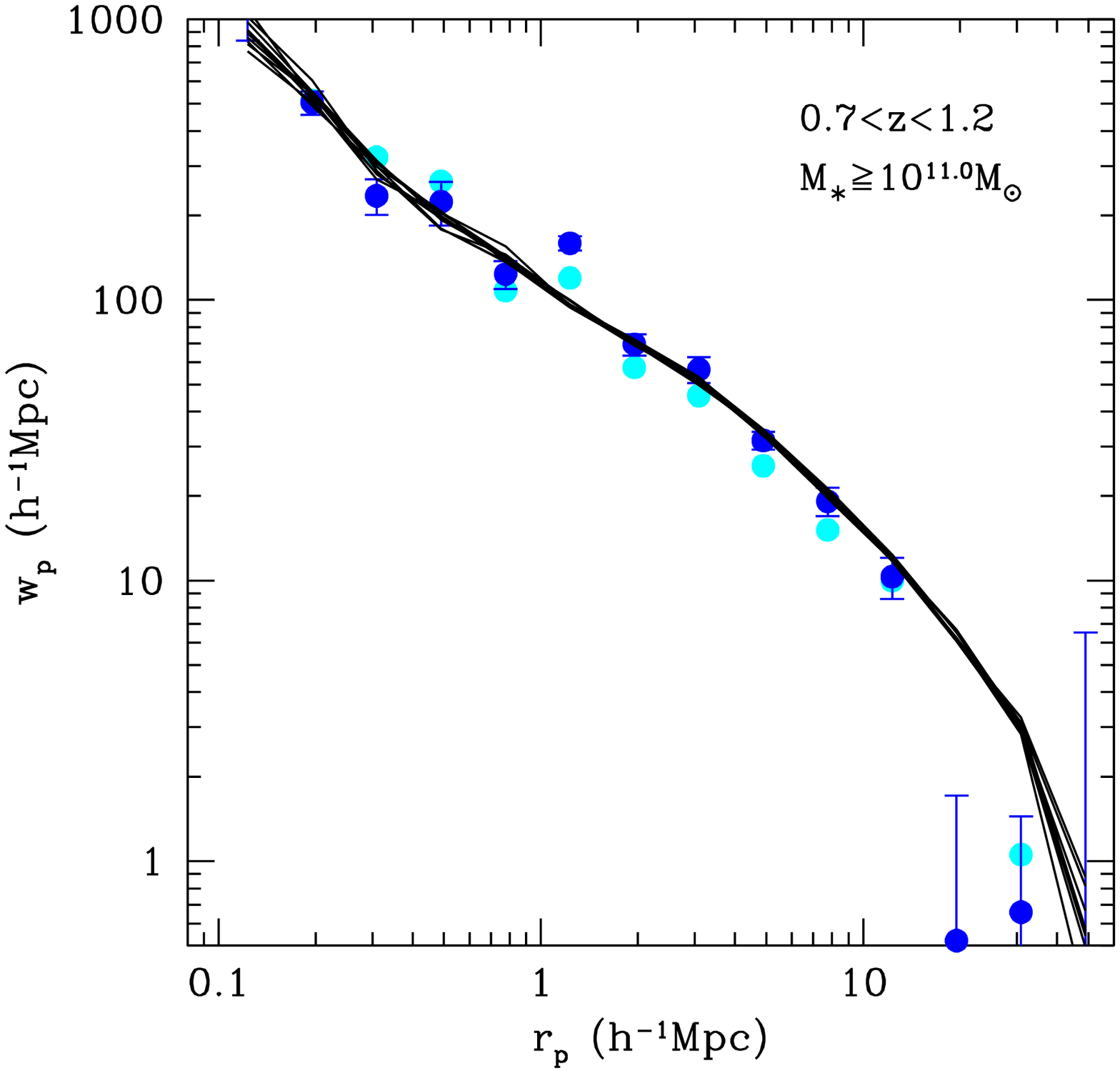} 
  \end{tabular}
 \caption{Top ten best-fitting halo occupation models (using clustering and abundance constraints) for the mock galaxy catalogs (described in Sec.~\ref{sec:mocks}) and the PRIMUS catalogs (described in Sec.~\ref{sec:PRIMUSCFs}).  The analytic models produce similar results, though with a 1-halo to 2-halo term transition that is too distinct (see Sec.~\ref{sec:mocks}). Each column shows results for different stellar mass thresholds, and upper (lower) rows show results at lower (higher) redshift.  Blue points indicate our fiducial measurements from all of the PRIMUS and DEEP2 fields, and cyan points indicate our measurements with the COSMOS field excluded.}
 \label{fig:HODexample}
\end{figure*}
\setlength{\tabcolsep}{6pt}

Following the procedure in the preceding section, we obtain best-fitting halo models with low values of $\chi^2$ relative to the clustering and abundances of galaxies in the five catalogs described in Section~\ref{sec:PRIMUSCFs}. 
In Figure~\ref{fig:HODexample}, we show measured projected two-point correlation functions from M15 and best-fitting mock catalogs for galaxies with stellar masses $\mathrm{log}~M_\ast>10.5$ at low and high redshift ($z\sim0.5$ and 0.9). 
 
In general, we find good agreement, in that our halo occupation models reproduce the PRIMUS correlation functions from M15 and the cumulative number densities (i.e., abundances from the stellar mass functions of Moustakas et al.\ 2013) well. 
For galaxies with $\mathrm{log}~M_\ast>10.5$ (middle panels) and with $\mathrm{log}~M_\ast>11.0$ (right panels), the measured correlation function at $z\sim0.9$ is slightly higher than at $z\sim0.5$, and we infer a slightly higher halo mass for the $z\sim0.9$ sample.  
The small-scale correlation functions (one-halo terms) are slightly different as well, indicating different satellite HOD parameters. 

\begin{figure}
   	\includegraphics[width=1.0\linewidth]{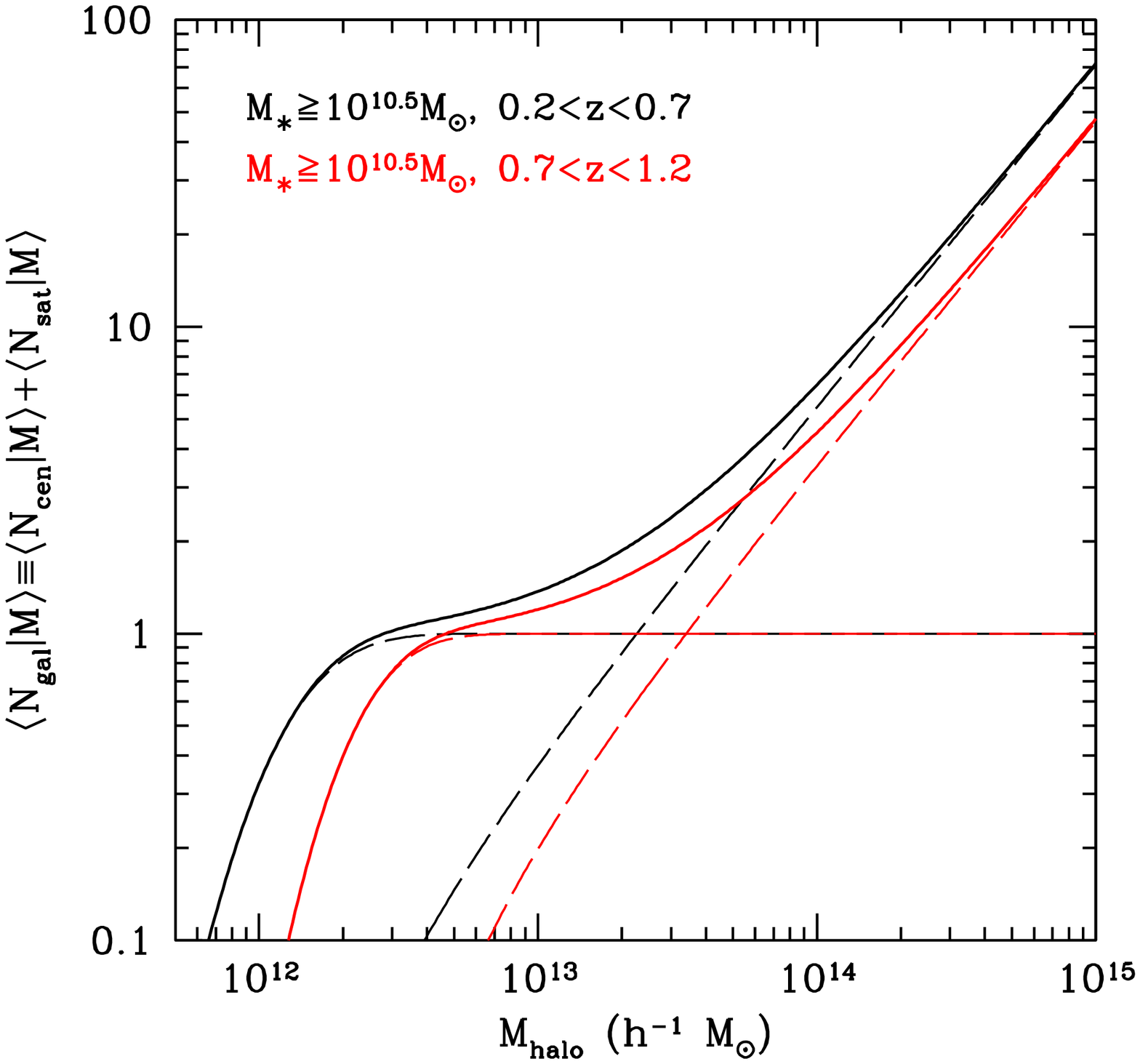} 
 	\caption{Mean halo occupation number, $\langle N|M\rangle$ for all, central, and satellite galaxies for best-fitting models based on analysis with mock catalogs of galaxies with $M_\ast\geq10^{10.5}M_\odot$. Low (high) redshift indicated by black (red) lines. 
        The shape of the mean HODs of the other three galaxy samples are similar. 
        The best-fit HOD parameters of all of the samples are presented in Tables~\ref{tab:HODresults} and \ref{tab:HODresults2}.}
 	\label{fig:NgalM}
\end{figure}

Next, in Figure~\ref{fig:NgalM} we show the mean of the halo occupation distributions, $\langle N_\mathrm{gal}\rangle$ (see Eqns.~\ref{eq:NcenM}-\ref{eq:NgalM}), for the best-fitting models for the two samples with galaxies of mass $M_\ast\geq10^{10.5}M_\odot$ as an example. 
Although the shapes of the distributions are similar, there are notable differences: in addition to the expected higher halo mass, the $z\sim0.9$ galaxies also have slightly higher $\sigma_{\mathrm{log}M}$ and slightly lower $M_1/M_\mathrm{min}$ parameter.

\subsection{HOD Results: inferred model parameters}

\begin{table*}
\caption{Halo Occupation Distribution Results: Mock Catalog Analysis}
 \centering
 \begin{tabular}{ l | c c c c c c }
   \hline
   sample & $\mathrm{log}M_\mathrm{min}$ & $\langle \mathrm{log}M_h\rangle$ & $\sigma_{\mathrm{log}M}$ & $M_1/M_\mathrm{min}$ & $\alpha$ & $f_\mathrm{sat}$ \\
   \hline 
   M1 & 11.70$\pm$0.05 & 12.00$\pm$0.10 & 0.20$\pm$0.10 & 17$\pm$1 & 1.05$\pm$0.05 & 0.22$\pm$0.02 \\
   M2 & 12.10$\pm$0.10 & 12.38$\pm$0.10 & 0.25$\pm$0.10 & 17$\pm$1 & 1.11$\pm$0.05 & 0.16$\pm$0.01 \\
   M3 & 12.60$\pm$0.10 & 12.85$\pm$0.10 & 0.20$\pm$0.10 & 16$\pm$1 & 1.09$\pm$0.05 & 0.11$\pm$0.01 \\
   M4 & 12.35$\pm$0.10 & 12.60$\pm$0.10 & 0.20$\pm$0.05 & 14$\pm$1 & 1.11$\pm$0.05 & 0.12$\pm$0.01 \\
   M5 & 12.65$\pm$0.10 & 12.87$\pm$0.10 & 0.20$\pm$0.10 & 15$\pm$1 & 1.12$\pm$0.05 & 0.08$\pm$0.01 \\
   \hline
  \end{tabular}
  \begin{list}{}{}
    \setlength{\itemsep}{0pt}
    \item HOD results for mock galaxy catalogs of the stellar mass and redshift-dependent (volume-limited) PRIMUS catalogs described in Table~\ref{tab:samples}.  $\langle \mathrm{log}M_h\rangle$ indicates the median halo masses; the mean halo masses are slightly higher (by $\approx0.1~\mathrm{dex}$ in most cases).  M1, M2, and M3 are at $z\sim0.5$ while M4 and M5 are at $z\sim0.9$.  M1 has the $\mathrm{log}M_\ast$ threshold 10.0, M2 and M4 have 10.5, and M3 and M5 have 11.0.
  \end{list}
 \label{tab:HODresults}
\end{table*}

\begin{table*}
\caption{Halo Occupation Distribution Results: Analytic Model Analysis}
 \centering
 \begin{tabular}{ l | c c c c c c }
   \hline
   sample & $\mathrm{log}M_\mathrm{min}$ & $\langle \mathrm{log}M_h\rangle$ & $\sigma_{\mathrm{log}M}$ & $M_1/M_\mathrm{min}$ & $\alpha$ & $f_\mathrm{sat}$ \\
   \hline 
   M1 & 11.70$\pm$0.05 & 12.00$\pm$0.10 & 0.20$\pm$0.10 & 18$\pm$1 & 1.09$\pm$0.05 & 0.21$\pm$0.02 \\
   M2 & 12.05$\pm$0.10 & 12.34$\pm$0.10 & 0.20$\pm$0.10 & 16$\pm$1 & 1.13$\pm$0.05 & 0.17$\pm$0.01 \\
   M3 & 12.60$\pm$0.10 & 12.85$\pm$0.10 & 0.20$\pm$0.10 & 15$\pm$1 & 1.10$\pm$0.05 & 0.11$\pm$0.01 \\
   M4 & 12.35$\pm$0.10 & 12.60$\pm$0.10 & 0.20$\pm$0.10 & 12$\pm$1 & 1.10$\pm$0.05 & 0.13$\pm$0.01 \\
   M5 & 12.65$\pm$0.10 & 12.87$\pm$0.10 & 0.20$\pm$0.10 & 12$\pm$1 & 1.03$\pm$0.05 & 0.10$\pm$0.01 \\
   \hline
  \end{tabular}
  \begin{list}{}{}
    \setlength{\itemsep}{0pt}
    \item Same as Table~\ref{tab:HODresults} but for the analytic halo model of galaxy clustering. 
  \end{list}
 \label{tab:HODresults2}
\end{table*}

We present the mass and redshift-dependent results of our analyses in Table~\ref{tab:HODresults} and in Figures~\ref{fig:MstMhrel1} and \ref{fig:fsatM}. 
%
The host halo masses at a given stellar mass are slightly different when only clustering constraints are used versus when those and abundances are used, which is our default method.  
For example, for the most massive galaxies at $z\sim0.5$ (M3), the measured clustering strength implies host halo masses of $M_h\geq10^{12.6}M_\odot/h$ while the number density favors a higher halo masses of $M_h\geq10^{12.9}M_\odot/h$. 
Conversely, for the $M_\ast\geq10^{10.5}M_\odot$ galaxies at $z\sim0.9$ (M4), the clustering strength implies host halo masses of $M_h\geq10^{12.40}M_\odot/h$ while including the number density results in a slightly lower value ($10^{12.25}$), though this may be due to anomalously strong clustering in COSMOS. 
We refer the reader to Appendix~\ref{app:withoutcosmos} for more details. 

We obtain very similar results with the analytic halo models (see Table~\ref{tab:HODresults2}) versus the mock galaxy catalogs. 
The halo masses are nearly identical, while the $M_1/M_\mathrm{min}$ values are slightly lower than for the mocks. 
We have tested the issue of the 1-halo to 2-halo transition region by artificially inflating the errors of the clustering measurements on those scales ($0.5<r_p<2~h^{-1}\mathrm{Mpc}$), and we obtained almost the same HOD parameters (within 1-$\sigma$), although we obtain slightly higher values (by a few percent) of the $\alpha$ parameter for the satellite galaxy HOD for some of the samples.

Note that the samples of galaxies selected at or above a particular stellar mass have number densities that evolve, so that the high-$z$ galaxies are not necessarily progenitors of their low-$z$ counterparts (Tojeiro et al.\ 2012; Leja et al.\ 2013). For this reason, some authors have rank ordered by luminosity or stellar mass and performed clustering analyses at a given number density (e.g., H.\ Guo et al.\ 2013; S14). 
However, for our range of stellar masses and redshifts, we obtain qualitatively similar results for mass- and number density-selected samples. (Note that the number densities of M2 and M4 and of M3 and M5 are very similar in Table~\ref{tab:samples}.)

\subsection{HOD Results: stellar mass dependence}

\begin{figure}
   \includegraphics[width=1.0\linewidth]{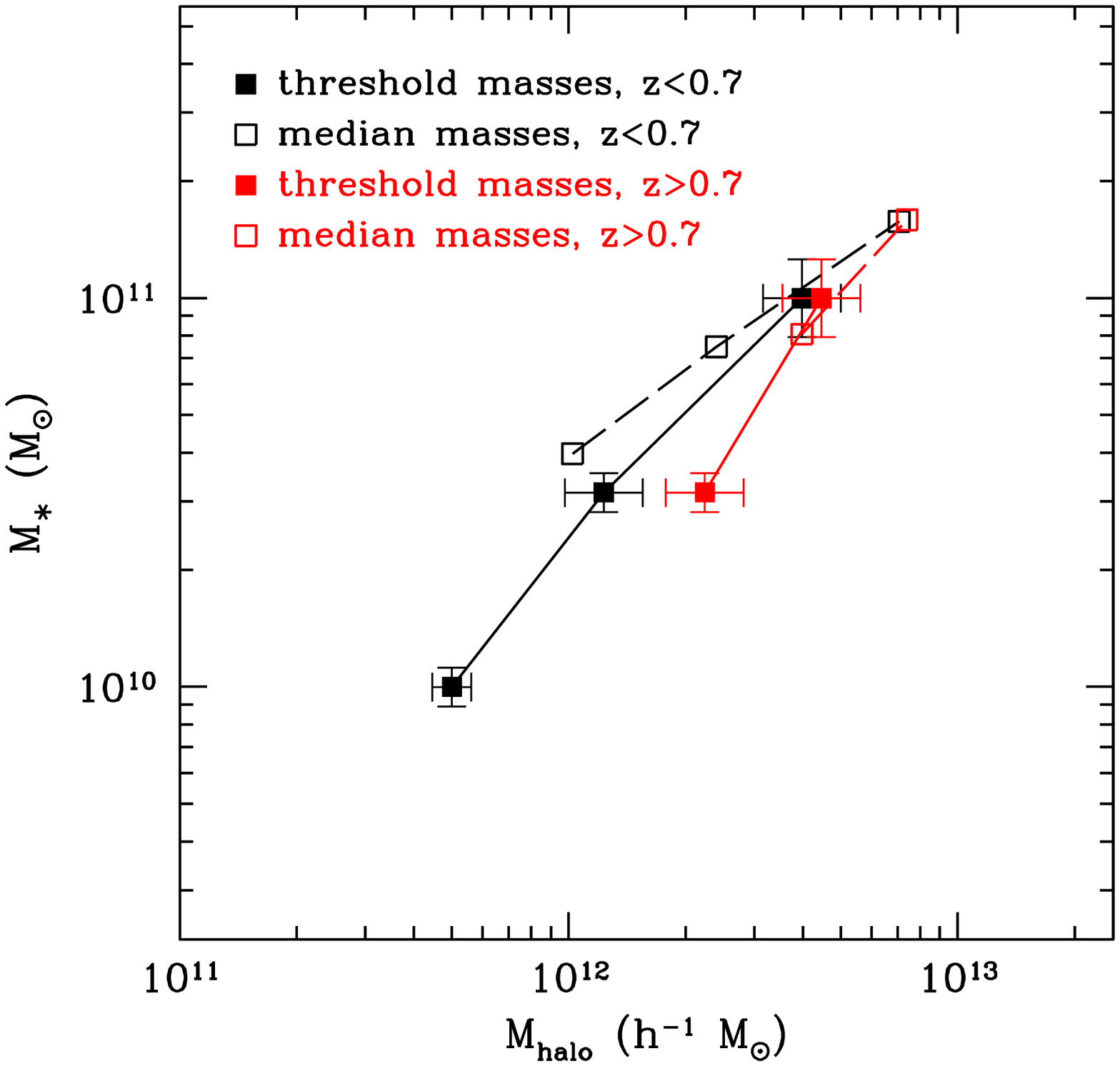} 
   \caption{Stellar mass-halo mass relation as a function of halo mass obtained from the analyses of HOD-based mock catalogs (see Table~\ref{tab:HODresults}). 
            Results are shown for mass thresholds (solid points with error bars) and median masses (open points with error bars omitted, for clarity).  
            Low (high) redshift results are indicated by black (red) points.}
\label{fig:MstMhrel1}
\end{figure}

First, we show the stellar mass-halo mass relation (SHMR) for both redshift ranges ($0.2<z<0.7$ and $0.7<z<1.2$) in Figure~\ref{fig:MstMhrel1}.  
Here and in what follows, we refer to the SHMR as the mean and rms of galaxy stellar mass as a function of halo mass, which quantify the distribution $p(M_\ast|M)$ (e.g., Behroozi et al.\ 2010; Leauthaud et al.\ 2012).

Two sets of relations are shown, such that threshold stellar masses ($M_\ast\geq10^{10.0}$, $10^{10.5}$, and $10^{11.0}$) are associated with threshold halo masses, and median stellar masses of galaxies in each sample are associated with a median halo mass. 
In both cases, the results are for best-fitting mock catalog-based models that minimized $\chi^2$ (\ref{eq:chi2}). 
As expected, we find a strong correlation between the masses at both redshifts, implying a steep slope in the SHMR over this mass range. 

\begin{figure}
   	\includegraphics[width=1.0\linewidth]{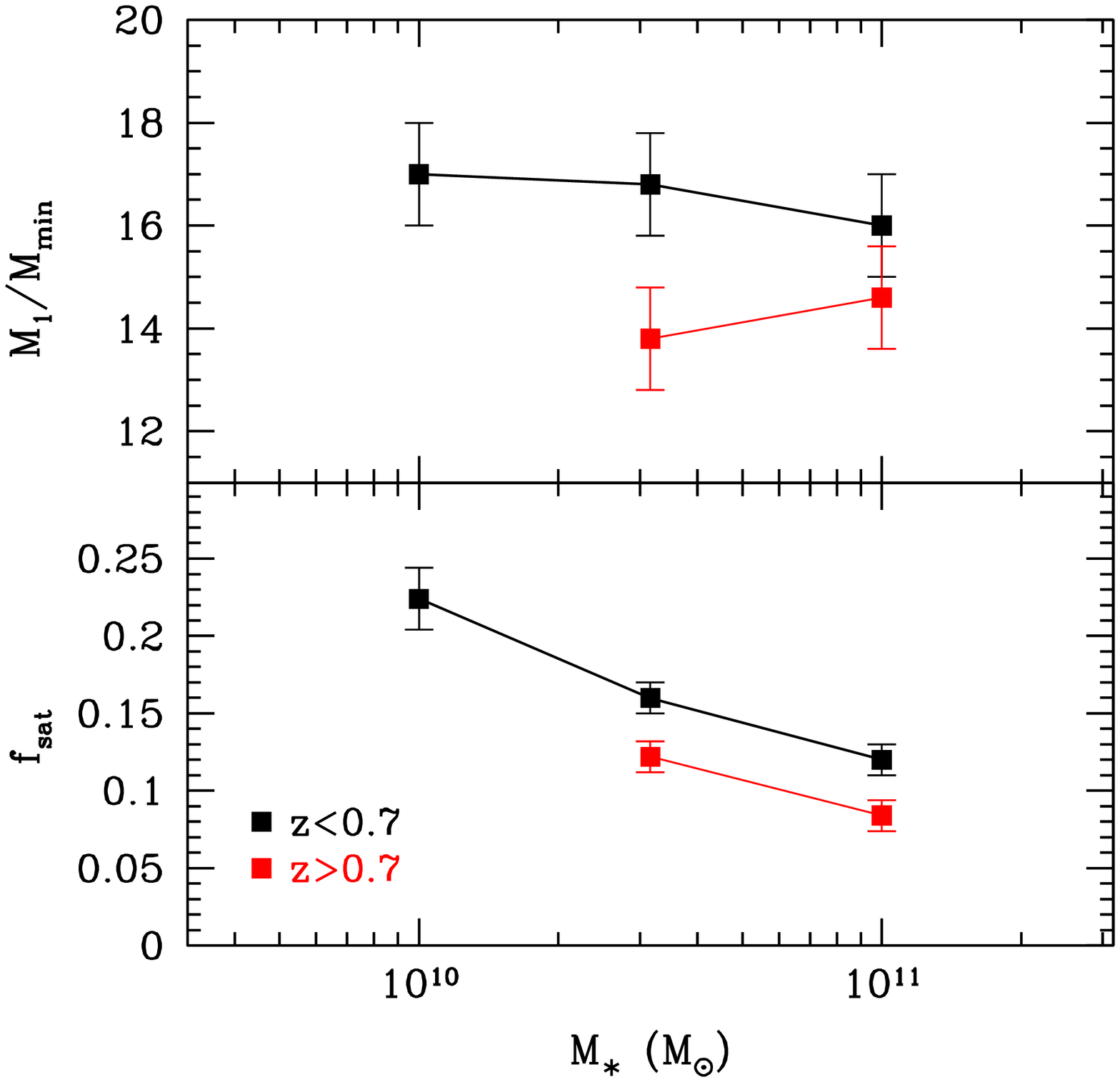} 
 	\caption{$M_1/M_\mathrm{min}$ parameter (upper panel) and satellite fraction (lower panel) as a function of stellar mass threshold obtained from the analyses of HOD-based mock catalogs. 
        Low (high) redshift results are indicated by black (red) points, and stellar mass error bars are omitted for clarity.}
 	\label{fig:fsatM}
\end{figure}

Second, we show the satellite galaxy halo occupation parameters, $M_1/M_\mathrm{min}$ and $f_\mathrm{sat}$, in Figure~\ref{fig:fsatM}.  The satellite fraction is simply defined by the following: 
\begin{equation}
 f_\mathrm{sat}(M_\ast) = \frac{
   \int_{M_\mathrm{min}} dM\,(dn/dM)\,\langle N_\mathrm{sat}|M\rangle}{
   \int_{M_\mathrm{min}} dM\,(dn/dM)\,(\langle N_\mathrm{cen}|M\rangle +
     \langle N_\mathrm{sat}|M\rangle)} .
 \label{eq:fsatM}
\end{equation}
We find that $M_1/M_\mathrm{min}$ is roughly constant with stellar mass, perhaps with some indication of an anticorrelation at lower redshift, consistent with H.\ Guo et al.\ (2014) and Zehavi et al.\ (2011). 
The satellite fraction decreases rapidly from $\approx25\%$ at low stellar masses to $\approx10\%$ at high masses, consistent with van den Bosch et al.\ (2007) at lower redshift.  
Most massive galaxies are centrals in massive haloes, and they are typically surrounded by less massive satellites. (We discuss these issues further in Section~\ref{sec:implications}.)

The values of $\sigma_{\mathrm{log}M}$ are not strongly constrained, but we find that they are approximately constant with mass and redshift, usually near a value of $\approx0.20$.  They are consistent with the satellite kinematics analysis of More et al.\ (2011), the clustering/lensing analysis of Cacciato et al.\ (2009), 
and the high-mass galaxy constraints of Shankar et al.\ (2014), 
while Zheng et al.\ (2007) and Leauthaud et al.\ (2012) obtained slightly higher values of $\approx0.3$ and $\approx0.25$. 
Note that the $\sigma_{\mathrm{log}M}$ parameter contains both a measurement error due to the stellar mass measure and redshift error and the intrinsic scatter. Moreover, different modeling frameworks (e.g., HOD, SHAM, as well as group catalogs and satellite kinematics) do not necessarily infer the same quantity, highlighting the difficulty of studying it precisely (see Leauthaud et al.\ 2011). 

\subsection{HOD Results: redshift evolution}\label{sec:HODz}

We find that the stellar mass-halo mass relation of central galaxies appears to evolve with redshift, with a significance of 2-3 $\sigma$. 
In particular, at stellar masses of $3\times10^{10}M_\odot$, the SHMR evolves such that a given galaxy stellar mass translates to a more massive halo at higher redshift.  
This is approximately consistent with other studies in the literature, and we compare and discuss them in Section~\ref{sec:MsMh}. 
%

The $M_1/M_\mathrm{min}$ parameter also evolves, with a decrease at higher redshift of 2.5 and 3.5-$\sigma$ significance for the mock catalogs and analytic model, respectively, and $f_\mathrm{sat}$ evolves over this range as well. 
Although haloes with a lower value of $M_1/M_\mathrm{min}$ will host \textit{more} satellites, note that a given stellar mass here does not translate to the same halo mass or number density over this redshift range, as discussed in the previous section. 
Haloes at fixed mass correspond to rarer density peaks at higher redshift and have lower number densities and fewer satellites these earlier epochs.  
We discuss implications of our results for satellite abundances and their formation and destruction in Section~\ref{sec:sats}. 
The other HOD parameters, $\alpha$ and $\sigma_{\mathrm{log}M}$, are approximately constant with redshift.

In the next section, we interpret these results and discuss their implications for central and satellite galaxy evolution in the context of the halo model.

%

\section{Discussion: Implications of the Results}\label{sec:implications}

\subsection{Stellar Mass-Halo Mass Relation}\label{sec:MsMh}

\begin{figure}[h!]
   \includegraphics[width=1.0\linewidth]{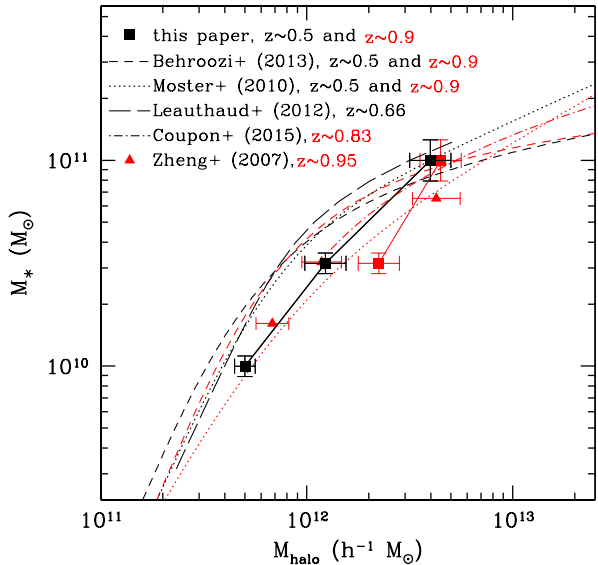}
   \caption{Stellar mass-halo mass relations obtained in this paper at $z=0.5$ and 0.9 (as in Fig.~\ref{fig:MstMhrel1}) compared to others in the literature, where black and red lines indicate lower- and higher-redshift results, respectively.  
   Results are shown for Moster et al.\ (2010; dotted lines at $z=0.5$ and 0.9); Behroozi et al.\ (2013; dashed lines at $z=0.5$ and 0.9), Leauthaud et al.\ (2012; long-dashed line at $z=0.66$), Coupon et al.\ (2015; dot-dashed line at $z=0.83$), all of which involve fitting relations to binned data, and Zheng et al.\ (2007; triangle points at $z\sim0.95$) based on luminosity-binned samples. 
   }
\label{fig:MstMhrel2}
\end{figure}

We now compare our mean stellar-to-halo mass relation (SHMR) for threshold masses to other studies in the literature in Figure~\ref{fig:MstMhrel2}. 
In particular, we compare our models of $\langle M|M_\ast\rangle$ to those of Moster et al.\ (2010) and Behroozi et al.\ (2010), 
which involve statistically determining the SHMR with a given shape and observed abundances (SMFs) as a function of stellar mass and redshift and stellar mass-dependent clustering at $z\sim0$. 
We also compare to Leauthaud et al.\ (2012), who fit to number densities, galaxy-galaxy lensing, and angular clustering in COSMOS, and to Coupon et al.\ (2015), who fit to number densities, lensing, and clustering in CFHTLenS. 
Necessary adjustments have been made to their quoted results when different IMFs or Hubble constant conventions were assumed.  
Other relevant results include luminosity dependent clustering analyses with VIPERS (de la Torre et al.\ 2013) and the COSMOS and CFHTLenS results for red and blue galaxies (Tinker et al.\ 2013; Hudson et al.\ 2015), 
though as these studies do not constrain the exact relation being studied here, we cannot compare our results directly. 
Within the 1-$\sigma$ error bars, our $z\sim0.5$ results are consistent with most of these studies, though there are some notable differences: for example, we obtain larger halo masses than Leauthaud et al.\ for galaxies with $M_\ast\sim10^{10.5}M_\odot$.  

At lower stellar masses, our results suggest a slightly stronger redshift evolution than obtained by others. 
For example, at $M_\ast=10^{10.5}M_\odot$, we estimate an increase in halo mass by 0.25-$0.30\pm0.10$~dex from $z\sim0.5$ to $z\sim0.9$, while the models of Moster et al.\ and Behroozi et al., which are not constrained by high-redshift clustering, predict a mass increase of no more than 0.1~dex over this redshift range---less than half as much evolution as we find.  
Our estimated SHMR evolution is not statistically significant ($0.15\pm0.10$~dex) when the COSMOS field is excluded (Table~\ref{tab:HODresults3}); however, the analysis of Leauthaud et al.\ finds less evolution within COSMOS based on the mean stellar and halo masses.  In another lensing analysis, Hudson et al.\ (2015) find no SHMR evolution for blue galaxies and $\approx0.2$~dex evolution for red ones.


Because of our limited dynamic range, it is difficult to fit a function to our SHMR results.  
Common double power-law parameterizations include the following ones of Moster et al.\ (2010) and Behroozi et al.\ (2010), respectively: 
\begin{equation}
 \frac{M_\ast}{M_h} = 2\biggl(\frac{M_\ast}{M_h}\biggr)_0 \biggl[ \biggl(\frac{M_h}{M_1}\biggr)^{-\beta} + \biggl(\frac{M_h}{M_1}\biggr)^\gamma \biggr]^{-1} 
 \label{eq:M10MsMhrel}
\end{equation}
\begin{eqnarray}
\mathrm{log}(M_h(M_\ast)) &=& \mathrm{log}(M_1) + \beta\mathrm{log}(\frac{M_\ast}{M_{\ast,0}}) \nonumber \\
 &+& \frac{(M_\ast/M_{\ast,0})^\delta}{1+(M_\ast/M_{\ast,0})^{-\gamma}} -\frac{1}{2},
 \label{eq:B10MsMhrel}
\end{eqnarray}
where $\beta$ and $\delta$ quantify the low- and high-mass slopes. 
Moster et al.\ obtain low- and high-mass slopes of approximately 2.2 and 0.5, respectively, while Behroozi et al.\ obtain slopes of approximately 2.6 and 0.3. 
It is possible to measure the relation's approximate slope to our model results as well, though the precise transition from the low- to high-mass regime is not clear.  
For galaxies at $z\sim0.5$ and using the threshold masses, if $M_\ast\propto M_h^\beta$, then $\beta\approx 1.3$ at $M_h\sim10^{12}M_\odot/h$ and $\beta\approx 1.0$ at $M_h\sim10^{12.6}M_\odot/h$.  When the median masses are used, the slope is slightly shallower by $\approx0.1$.  
At $z\sim0.9$, the slope is slightly steeper than that of other models for halo masses less than $M_h\sim10^{12.6}M_\odot/h$.
At higher masses, for group and cluster haloes, the Moster et al.\ and Behroozi et al.\ models predict that the slope increases by about 20-25\% between $z\sim0.5$ and 0.9, though our model constraints are not sufficiently precise at such masses to determine whether we obtain similar evolution.
 
Rather than analyze this relation, some authors take the ratio, $f_\ast\equiv M_\ast/M_h$, which does not contain new information but more clearly indicates the peak or `pivot' mass scale (e.g., van den Bosch et al.\ 2007; Leauthaud et al.\ 2012). 
Based on our results in Table~\ref{tab:HODresults} and Figure~\ref{fig:MstMhrel1}, we find that $f_\ast$ peaks at a mass scale of $12.0<\mathrm{log}M_h<12.4$ at a value of 0.028, consistent with the literature and well below the universal baryon fraction of approximately $15.5\%$ (assuming $\Omega_b h^2=0.022$ and $\Omega_c h^2=0.120$ from Planck collaboration et al.\ 2014; see also Fukugita \& Peebles 2004). 
At $z\sim0.9$, the pivot (log) halo mass appears to shift upward to $\approx12.3$-12.6, slightly higher than studies in the literature, though a wider dynamic range with high completeness would be necessary to analyze this further and obtain more precise results. 


It is well known that the SHMR has a significant amount of scatter (More et al.\ 2011; Skibba et al.\ 2011; Rodr\'{i}guez-Puebla et al.\ 2013). 
The relation's scatter in our best-fit models, quantified by $\sigma_{\mathrm{log}M}$, is consistent with the literature but has large error bars.
Note that our central galaxy HOD assumes a lognormal distribution of stellar mass at fixed halo mass (Eqn.~\ref{eq:PcenM}); although this assumption is consistent with the current data available, it may not be entirely accurate.
Also note that SHMR studies in the literature do not all assume the some halo mass definition (see App.~\ref{app:Mdef}), though even for the same definition, halo occupation and semi-analytic models make a wide range of predictions for the SHMR at a given halo mass (Knebe et al.\ 2015).

Finally, mean or median SHMRs of central and satellite galaxies are similar when each object's halo mass (or circular velocity) at the approximate time of accretion is used, though we refer the reader to Rodr\'{i}guez-Puebla et al.\ (2012, 2013) and Watson \& Conroy (2013) for analyses of some differences between them. 
However, as a function of parent halo mass, satellites have surprising mass distributions and relations, with implications for their co-evolution with the subhalos that host them, and we turn to these issues in the following subsection.

\subsection{Satellite and Subhalo Abundances}\label{sec:sats}

In Section~\ref{sec:HODz}, we noted that the satellite HOD (described in Sec.~\ref{sec:HOD}), and in particular the $M_1/M_\mathrm{min}$ parameter (or $\mu$) appears to evolve, consistent with other studies (e.g., de la Torre et al.\ 2013).  
Combining our satellite HOD results with lower-redshift results from the SDSS (Zehavi et al.\ 2011) and higher-redshift results at $z\sim1$ (Zheng et al.\ 2007) and at $z\sim1.5$ (Wake et al.\ 2011; Martinez-Manso et al.\ 2014) imply a clear redshift dependence. 
For example, applying the function $M_1/M_\mathrm{min}(z)=\mu_0-\beta z$ to all of these results yields $\mu_0\approx19$-20 and $\beta\approx6$-7. 

\subsubsection{Satellite Mass Function}

The decreasing $M_1/M_\mathrm{min}$ with increasing redshift is related to the competition between merging and satellite destruction (see Conroy et al.\ 2006; Wetzel et al.\ 2009).
These results have implications for satellite abundances and fractions 
as a function of mass and redshift. One way to study this is to revisit the analysis of satellite galaxy and subhalo occupation distributions of Skibba et al.\ (2007); we refer the reader to that paper for more details. 

Firstly, note that although the \textit{most massive} satellite's mass or luminosity will scale with the host halo, the mean satellite mass or luminosity is nearly flat over many orders of magnitude of parent halo mass (see also Skibba et al.\ 2011; Paranjape \& Sheth 2012). 
Our analysis also predicts that not just the mean, but the shape of the satellite galaxy stellar mass function is approximately independent of halo mass. 

As we argued in Skibba et al.\ (2007) for galaxy luminosities, the power-law shape of the mean satellite HOD (\ref{eq:NsatM}) and the approximate exponential shape of the stellar-to-halo mass relation (except at low masses\footnote{Over a wide mass range, including masses well below the knee of the SMF, a double power-law shape is more accurate (Moster et al.\ 2010; Behroozi et al.\ 2010; see also Moustakas et al.\ 2013).}) implies that the satellite conditional SMF, $p(M_{\ast,\mathrm{sat}}|M)$, is approximately has a Schechter-like function. 
This is borne out in the analysis of central and satellite conditional stellar mass distributions in group catalogs by Yang et al.\ (2009). The evolving $\mu$ parameter with redshift only implies that the \textit{amplitude} of the satellite conditional SMF evolves, while the shape of it remains constant at least out to $z\sim0.9$. 
Therefore, satellite abundances or occupation numbers of haloes evolve, but low and high-redshift haloes have similar relative abundances of low vs high-mass satellites. 

\subsubsection{Intracluster Mass}



We can use our results to determine the approximate amount of stellar mass in a diffuse stellar halo or the intracluster medium, likely due to disrupted satellite galaxies. 
Such a calculation is made by comparing the halo mass fraction in subhalos to the stellar mass fraction in satellites (Skibba et al.\ 2007; White et al.\ 2007; Yang et al.\ 2009). 
We utilize our constraints on the stellar mass and redshift dependent halo occupation parameters, especially $M_1/M_\mathrm{min}$, and we compare the relative abundances of satellite galaxies to that of subhaloes estimated from the subhalo mass function (MF). 
From Giocoli et al.\ (2010), the subhalo MF can be expressed as the following: 
\begin{eqnarray}
 {{\rm d}N(m|M)\over {\rm d}m}\, {\rm d}m = N(z)\,
  \left({M\over 10^{12} h^{-1}M_\odot}\right)^{0.1}\nonumber \\
  \left({M\over m}\right)^{0.9}\,\mathrm{exp}\biggl[-\beta\biggl(\frac{m}{M}\biggr)^3\biggr]\,{{\rm d}m\over m}
 \label{eq:dNgiocoli}
\end{eqnarray} 
where $m$ is the subhalo mass at the redshift of interest, $N(z)\approx0.0148(1+z)^{1/2}$ and $\beta\approx12.2715$. 
If we neglect the exponential in (Eqn.~\ref{eq:dNgiocoli}), then 
\begin{eqnarray}
  N(\ge m|M) &=& \int_m {\rm d}m\, {{\rm d}N(m|M)\over {\rm d}m}\, {\rm d}m \\
   &\approx& \frac{N(z)}{0.9}\,
   \left({M\over 10^{12} h^{-1}M_\odot}\right)^{0.1}\,
   \left({M\over m}\right)^{0.9}. \nonumber
\end{eqnarray}
If we use $M_1$ to denote the value of $M$ at which the number of
subhaloes is unity, then the expression above implies that
\begin{equation}
\left({M_1\over m}\right) 
  \approx \frac{60.8}{(1+z)^{1/2}}\,\left({10^{12} h^{-1}M_\odot}\over m\right)^{0.1}
\end{equation}

Going a step further, using the subhalo MF in (Eqn.~\ref{eq:dNgiocoli} and neglecting the exponential\footnote{According to Giocoli et al.\ (2010), neglecting the exponential is not entirely accurate, but it is sufficiently accurate for the very approximate calculation here. The integral can be performed analytically with the exponential and yields incomplete gamma functions in the result.}), the mass fraction in subhaloes is approximately given by 
\begin{eqnarray}
 f_\mathrm{sub}(M) &=& \int_0^M {m\over M}\, {{\rm d}N(m|M)\over {\rm d}m}\, {\rm d}m \nonumber\\
  &=& 0.148(1+z)^{1/2}\,\left({M\over 10^{12} h^{-1}M_\odot}\right)^{0.1},
 \label{eq:massfrac}
\end{eqnarray}
where if stars only form in sufficiently massive objects, the lower limit to this integral may be greater than zero. 
The $\mu\equiv M_1/M_\mathrm{min}$ parameter of galaxies is as low as $\approx12$ at $z\sim1$, 
consistent with Martinez-Manso et al.\ (2015) and other studies, 
but for subhaloes, the ratio is much larger, implying that some mass has been stripped from satellites, presumably contributing to stellar mass in the intracluster light (ICL). 
Revisiting the calculation in Skibba et al.\ (2007) but with new clustering and subhalo constraints, we argue that the mass fraction in the intracluster medium is given by 
\begin{eqnarray}
 f_\mathrm{ICM}(M,M_\ast) &=& 1\,-\, \biggl(f_\mathrm{sub}(M,z)\frac{60.8(1+z)^{-1/2}}{\mu(z)}\biggr) \\
 &\times& \biggl(\frac{\langle N_\mathrm{sat}|M,M_\ast\rangle + \langle M_{\ast,\mathrm{cen}}|M\rangle/\langle M_{\ast,\mathrm{sat}}|M\rangle}{\langle N_\mathrm{sat}|M,M_\ast\rangle}\biggr), \nonumber
 \label{eq:fICL}
\end{eqnarray}
where the halo redshift dependencies cancel, implying that this fraction's evolution depends primarily on $\mu(z)$---that is, on how the halo occupation number of subhaloes relative to satellite galaxies evolves.  

The halo model predicts the ICL fraction to increase with host halo mass, in agreement with other analyses (Murante et al.\ 2007; Purcell et al.\ 2007), which also highlight uncertainties about the fate of `orphan' satellites over time (Conroy et al.\ 2007; Contini et al.\ 2014). 
For halo masses of $M\sim10^{14}h^{-1}M_\odot$ and galaxies with $M_\ast\geq10^{10.5}M_\odot$, we estimate that this fraction is approximately 5\% at $z\sim0.5$ and less than 2\% at $z\sim0.9$. For halo masses of $M\sim10^{15}h^{-1}M_\odot$ the ICL fraction is approximately three times larger, though the rarity of such massive haloes at high redshift makes this calculation even more uncertain. It appears that the stripping and disruption of satellites indeed contribute to the ICL over the 2.5~Gyr between these epochs, but additional analysis with more precise halo-model parameters is required to study this further. 



%

\section{Conclusions}\label{sec:conc}

In this paper we analyze the stellar mass and redshift dependent spatial clustering and abundances 
of galaxies in PRIMUS and DEEP2 at $0.2<z<1.2$ with $\Lambda$CDM halo models of galaxy clustering.  
In order to obtain robust results for model parameters, we perform the analysis with two independent sets of models: HOD-based mock galaxy catalogs and analytic models. 

We summarize our main conclusions as follows: 
\begin{itemize}
 \item Both sets of models are able to accurately reproduce the stellar mass and redshift dependent correlation functions and number densities.  The best-fit models yield consistent and robust halo occupation parameters for central and satellite galaxies. 
 \item The results for central galaxies constrain the redshift evolution of the mean stellar-to-halo mass relation (SHMR). For halo masses below $10^{12.6}h^{-1}M_\odot$, we find that the SHMR appears to evolve significantly, such that galaxies of a given stellar mass are associated with more massive haloes at higher redshifts. 
 We find that the ratio $M_\ast/M_\mathrm{halo}$ peaks at a value of 0.028 at the mass scale $12.0<\mathrm{log}(M_\mathrm{halo}/h^{-1}M_\odot)<12.4$.
 We do not obtain statistically significant constraints on the evolution of the scatter in the SHMR. 
 \item The satellite fraction increases rapidly with decreasing stellar mass and the $M_1/M_\mathrm{min}$ parameter, which quantifies the critical mass above which haloes host at least one satellite, is more redshift than mass dependent, with lower values ($\approx12$-15) at high redshift.  We use the HOD of satellites and subhaloes to estimate rough constraints on the mass fraction of disrupted satellites that contribute to the intracluster medium. 
\end{itemize}

Our joint analysis with analytic models and mock catalogs, which produce consistent results, demonstrates their robustness and the strength of our conclusions. 
Abundance-matching with numerical simulations and studies with hydrodynamic simulations are currently popular in the field, but analytic models remain an important tool as well. All models have some assumptions, uncertainties, and free parameters, and every model has advantages and shortcomings; therefore, it is important to utilize multiple types of models when possible. 
The advantages of analytic models include their speed and computational inexpensiveness and their ability to parameterize physical processes and galaxy-halo relations in a way that aids understanding their origins.

By comparing galaxy counterparts at different redshifts, one may constrain the extent to which galaxies of a given mass or number density grow by in situ star formation vis-\`{a}-vis mergers with neighboring galaxies over a given redshift range (Wake et al.\ 2008; Zehavi et al.\ 2012; Lackner et al.\ 2012). 
For example, Wake et al.\ (2008) utilized a similar analytic halo model at $0.19<z<0.55$ as the one used here to estimate the central galaxy merger rate of luminous red galaxies. For the galaxy samples used here, however, the clustering measurements and abundances are consistent with passive evolution, suggesting a very low merger rate, though the uncertainties are too large to make quantitative estimates (S14). Nevertheless, Seo et al.\ (2009) argue that a shoulder in the HOD ($M_1\gg M_\mathrm{min}$) implies that a galaxy population has not undergone passive evolution, and this remains an important issue for future research. 
In addition, analysis of the clustering of star-forming and quiescent galaxies out to $z\sim1$ in PRIMUS and DEEP2 would greatly benefit the field and would constrain the quenched fractions of central and satellite galaxies as a function of mass and redshift, which would be important for distinguishing between competing models (Cohn \& White 2013). 
%

Furthermore, as described in Section~\ref{sec:model}, throughout this paper we have assumed that the evolving spatial distributions of galaxies and their correlations with the dark matter haloes are primarily determined by the mass of the haloes. 
This implies an implicit assumption that `galaxy assembly bias,' where galaxies' distributions depend on the assembly history of systems at fixed halo mass, is such a small effect as to be negligible for the analysis. 
Some recent studies in the literature (Zentner et al.\ 2014; Hearin et al.\ 2014) argue that this assumption may be incorrect and therefore inferred model parameters and galaxy-halo correlations, such as the ones obtained in this paper, may be biased.  
Addressing this question is beyond the scope of this paper, but it too will be the focus of subsequent research. 

Finally, this paper is complementary to other current and upcoming work. 
In particular, stellar mass and SFR dependent clustering measurements in PRIMUS will be presented in Mendez et al.\ (in prep.; M15), luminosity and color dependent cross-correlation functions are presented in Bray et al.\ (2015), and the environmental dependence of stellar mass functions are in Hahn et al.\ (2015).  In addition, complementary modeling of SFR dependent clustering will be presented in Watson et al.\ (in prep.) extending the work in Watson et al.\ (2015), and details about the models used for constructing mock galaxy and group catalogs will be in Skibba (in prep.).

\section*{Acknowledgments}
RAS and ALC acknowledge support from the NSF CAREER award AST-1055081. 
TM is supported by UNAM-DGAPA PAPIIT IN110209 and CONACyT Grant \#83564. 
We thank the Aspen Center for Physics, which has been supported by NSF grant PHYS-1066293, for hospitality during the summer 2014 workshop on the ``Galaxy-Halo Connection Across Cosmic Time," where some of this work was completed. 
We acknowledge Peter Behroozi, Andrew Hearin, Alexie Leauthaud, Ravi Sheth, and Doug Watson for valuable discussions about our models and results, 
and we thank Idit Zehavi and Zheng Zheng for helpful comments on a previous draft. 
We also thank the anonymous referee for valuable recommendations that improved the quality of the paper. 
We acknowledge Scott Burles and Kenneth Wong for their contributions to the PRIMUS project. 
Funding for PRIMUS has been provided by NSF grants AST-0607701, 0908246, 0908442, 0908354, and NASA grant 08-ADP08-0019.




\bibliographystyle{apj}

\appendix

\renewcommand{\thefigure}{\Alph{appfig}\arabic{figure}}
\setcounter{appfig}{1}
\renewcommand{\thefigure}{\Alph{apptab}\arabic{figure}}
\setcounter{apptab}{1}

\section{Halo Masses and Halo Finding Algorithms}\label{app:Mdef}

As noted in Section~\ref{sec:model}, throughout this paper we use halo masses and radii 
defined using a virial overdensity 200 times the mean density of the Universe for the 
analytic model or defined using the evolving virial overdensity specified by Bryan \& 
Norman (1998) for the mock catalogs.  These and other halo definitions typically used 
in the literature, including spherical overdensities (SO), Friends-of-Friends (FoF), and phase-space ones, will have only minor effects on the kinds of quantitative results presented here, 
and the qualitative trends will remain unchanged. 

In general, SO halo finders tend to impose a more spherical geometry on the resulting systems, while FOF sometimes links neighboring objects via tenuous bridges of particles.  In either case, the choice of virial overdensity and halo membership can be important. 
Dynamically unrelaxed haloes, which are common, as well as poorly resolved haloes have poorly estimated or biased masses, concentrations, and other affected parameters (e.g., Skibba \& Macci\`{o} 2011).  
In addition, some authors adopt different circular velocity definitions, such as $V_\mathrm{max}$ and $V_\mathrm{peak}$, which may affect some model-dependent results (Reddick et al.\ 2013; Behroozi et al.\ 2014). 

A detailed analysis of halo definitions is beyond the scope of this paper but will be important for advancing the fields of galaxy formation and large-scale structure formation. 
For more studies of the effects of halo definitions, we refer the reader to Knebe et al.\ (2011), Zemp (2014), Klypin et al.\ (2014); and More, Diemer \& Kravtsov (2015); for studies of subhalo definitions, we refer the reader to Onions et al.\ (2012) and Pujol et al.\ (2014).



\section{Halo Mass Function and Bias}\label{app:modelstuff}

\subsection{Matter Power Spectrum}

The linear matter power spectrum is described by
\begin{equation}
  P_\mathrm{mm}^\mathrm{lin}(k,z) \,\propto\, T^2(k)\,k^{n_s}\,D^2(z)/D^2(0)
\end{equation}
where we use the Smith et al.\ (2003) matter power spectrum model and the Eisenstein \& Hu (1998) transfer function, which assumes a CMB temperature of $2.725~K$.  
$D(z)$ is the linear growth factor and we have set the spectral index $n_s=1$ (see e.g., Dodelson 2003). 

The mass variance is described by
\begin{equation}
  \sigma^2(M)\,=\,\frac{1}{2\pi^2} \int_0^\infty P_\mathrm{mm}^\mathrm{lin}(k,z)\, \widehat{W}^2(kR)\, k^3\, \frac{\mathrm{d}k}{k}
\end{equation}
which is set such that it is equal to $\sigma_8^2$ for $R=8h^{-1}\mathrm{Mpc}$ at $z=0$. $\widehat{W}$ is the Fourier transform of the top-hat window function of radius $R$:
\begin{equation}
  \widehat{W}(kR) = \frac{3[\mathrm{sin}(kR) - kR\mathrm{cos}(kR)]}{(kR)^3}
\end{equation} 

\subsection{Halo Mass Function}

We assume the Tinker et al.\ (2008) halo mass function\footnote{This is a modification from our previous models, in which we used the Sheth, Mo, \& Tormen (2001) mass function.} in which haloes are identified with a SO algorithm with $\Delta=200$ and $\bar \rho_m(z)=\Omega_m(z)\rho_\mathrm{crit}(z)=\bar \rho_m(0)(1+z)^3$.  
Halo abundances are assumed to follow a universal function in terms of the mass fraction of matter in peaks of height $\nu=\delta_c/\sigma(M,z)$, where $\delta_c=0.15(12\pi)^{2/3}\approx1.686$\footnote{Note that van den Bosch et al.\ (2012) use a slightly different definition that includes a factor of $[\Omega_m(z)]^{0.0055}/D(z)$.}. 
The halo abundances are well described by the following functional form:
\begin{equation}
  \frac{dn}{dM}\,=\,f(\sigma)\frac{\bar \rho_m}{M}\frac{d\mathrm{ln}\sigma^{-1}}{dM}
\end{equation}
where
\begin{equation}
 f(\sigma)\,=\,A\biggl[\biggl(\frac{\sigma}{b}\biggr)^{-\alpha}+1\biggr]e^{-c/\sigma^2}
\label{eq:fsigma}
\end{equation}
and the parameters $A$, $a$, and $b$ are redshift dependent and constrained by numerical simulations that are described in Tinker et al. (2008).

We find that this analytic function yields halo abundances as a function of mass and redshift that are approximately consistent with Bolshoi (Klypin et al.\ 2011) Rockstar (Behroozi et al.\ 2013) catalogs in which masses are defined with an evolving virial overdensity (Bryan \& Norman 1998).
Note that the halo finder used by Tinker et al.\ allows for overlapping haloes. 

The characteristic mass of the halo mass function is defined at the scale at which a typical peak ($\nu=1$) collapses at a given redshift: $\sigma(M^*,z)=\sigma(M^*,0)D(z)=\delta_c(z)$. 
$M^*(z)$ decreases with increasing redshift, from a few times $10^{12}h^{-1}M_\odot$ at $z\sim0$ to below $10^7h^{-1}M_\odot$ at $z\sim6$ (Mo \& White 2002).  

\subsection{Halo Bias Function}

We adopt the Tinker et al.\ (2010) halo bias function with $\Delta=200$, 
in which the bias is expressed in the following flexible form
\begin{equation}
 b(\nu)\,=\, 1-A\frac{\nu^a}{\nu^a+\delta_c^a}+B\nu^b+C\nu^c,
 \label{eq:bT10}
\end{equation}
where the parameters $A$, $a$, and $C$ are a function of $\Delta$. 
This bias function is forced to obey the relation
\begin{equation}
 \int d\nu~b(\nu)f(\nu)=1,
\end{equation}
where $f(\nu)$ is the halo mass function [such that $f(\sigma)=\nu f(\nu)$ using the notation in (\ref{eq:fsigma})]. 
The Tinker et al.\ (2010) departs from the bias model of Sheth, Mo, \& Tormen (2001) at very low and very high values of $\nu$. 
We find that the alternate bias function of Tinker et al.\ (2010) using the peak-background split yields similar clustering predictions as (\ref{eq:bT10}).

Scale-dependent bias (at fixed mass) and halo exclusion effects could slightly affect the transition from the 1-halo to 2-halo terms and the inferred satellite occupation parameters, and an analysis of models of these effects is the subject of ongoing work.  
one approach is that of van den Bosch et al.\ (2013) and Tinker et al.\ (2005): 
\begin{equation}
  P_{2h}(k) \,=\, \Biggl[ \int dM \,\frac{dn(M)}{dM}\,
                 \langle N_{\mathrm {cen}}|M\rangle
          \frac{1\,+\,
          \langle N_{\mathrm {sat}}|M\rangle\,
	  u_{\mathrm {gal}}(k|M)}{\bar n_{\mathrm {gal}}}\Biggr]^2 \,
	  Q(k|M), \label{eq:Pk2hvdB} 
\end{equation}
where 
\begin{equation}
  Q(k|M) \equiv 4\pi \int_{r_\mathrm{min}(M)} dr r^2 \frac{\mathrm{sin}(kr)}{kr}\, 
    [  1+b_h(M_1)b_h(M_2)\,\zeta(r)\xi_{mm}^{\mathrm {lin}}(r)  ] 
\label{eq:Q}
\end{equation}
$r_\mathrm{min}(M)$ is the cutoff that accounts for halo exclusion, and $\zeta(r)$ quantifies the scale dependence of bias based on some fitting functions. 

Considering that the matter power spectrum is modeled in Fourier space, we have opted to model galaxy clustering in Fourier space as well. Therefore, implementing the formulation above for our analytic models would require an additional integral and would be more computationally expensive. 
Another promising approach of scale-dependent bias is the excursion set analysis of Musso et al.\ (2012) and Paranjape et al.\ (2013) (cf., Smith et al.\ 2007; Desjacques et al.\ 2010). 
This approach predicts the following form for the bias factors: 
\begin{equation}
 b_n = \biggl(\frac{S_\times}{S_0}\biggr)^n \sum_{r=0}^n {n \choose r} b_{nr} \epsilon_\times^r
\end{equation}
where we are interested only in the $n=1$ case, and 
\begin{eqnarray}
 S_\times &=& \int d\mathrm{ln}k\,\Delta^2(k)W(kR)W(kR_0) \nonumber\\
 \epsilon_\times &=& 2~d\mathrm{ln}S_\times/d\mathrm{ln}s
\end{eqnarray}
are cross-correlations between the mass overdensity field smoothed on the large scale $R_0$ and the Lagrangian scale of the halo $R$.

More work is required to accurately model scale-dependent bias and halo exclusion. 
We also find degeneracies between the parameters of these models and between them and the treatment of halo profile truncation (see the following section).

\subsection{Density Profile}

For the halo density profiles, we assume spherical symmetry and adopt a Navarro, Frenk \& White (1997; NFW) profile:
\begin{equation}
  \rho(r|M) \,=\, \frac{\rho_s}{(r/r_s)(1+r/r_s)^2}
\end{equation}
an NFW profile is not a sufficiently accurate description of the small-scale distribution of satellite galaxies (Watson et al.\ 2012; Piscionere et al.\ 2014), but this affects the clustering at smaller separations than we can accurately probe.

The Fourier transform of the NFW density profile is the following: 
\begin{equation}\label{eq:ukM}
  u(k|M) \,=\, \int_0^{r_\mathrm{vir}}\, dr\,4\pi r^2 \frac{\mathrm{sin}(kr)}{kr} \frac{\rho(r|M)}{M}
\end{equation}
For an NFW profile, the integral can be computed analytically (see also Scoccimarro et al.\ 2001; Cooray \& Sheth 2002), and the solution to the indefinite integral is 
\begin{equation}\label{eq:ukMintd}
  u(k|M) \,=\, f(c) \Biggl\{ \mathrm{sin}(kr_s) \mathrm{Si}[k(r+r_s)] \,+\, \mathrm{cos}(kr_s) \mathrm{Ci}[k(r+r_s)] \,-\, \frac{\mathrm{sin}(kr)}{r+r_s} \Biggr\}
\end{equation}
where $\mathrm{Ci}(x)$ and $\mathrm{Si}(x)$ are the cosine and sine integrals, respectively. 
$c\equiv r_\mathrm{vir}/r_s$ and the virial radius is defined to be $r_{200m}$. 
The expression in (\ref{eq:ukMintd}) is evaluated from $r=0$ out to $r=r_\mathrm{vir}$, but if the profile is not truncated at the virial radius, then the above reduces to $f(c) \mathrm{sin}(kr_s) \pi/2$ as $r\rightarrow\infty$ (see Sheth et al.\ 2001 for a discussion of halo truncation).  Other profiles such as a Moore et al.\ (1999) or Einasto (1965) profile may be substituted in (\ref{eq:ukM}), but they must be integrated numerically. 

We assume the redshift-dependent concentration-mass relation of Mu\~{n}oz-Cuartas et al.\ (2011), and in our mock catalogs, we include the scatter in the relation. 
The relation is fitted for dynamically relaxed haloes identified with a spherical overdensity algorithm and has the form
\begin{equation}
 \mathrm{log}(c)=a(z)\mathrm{log}(M_\mathrm{vir}/[h^{-1}M_\odot])+b(z)
\end{equation}
We obtain similar clustering predictions using the concentration mass relation of Macci\`{o} et al.\ (2008).

The concentration and velocity dispersion of galaxies and subhaloes is different than that of dark matter particles (e.g., Munari et al.\ 2013; Old et al.\ 2013), and we attempt to account for this effect as well.  
Note that studies in the literature currently disagree about the extent to which satellite galaxies are less concentrated than dark matter (see also Yang et al.\ 2005; Hansen et al.\ 2005).

\section{Cosmic Variance and the Effect of the COSMOS Field}\label{app:withoutcosmos}

In Section~7.5 of S14, we discussed the effects of `cosmic variance,' or field-to-field fluctuations, and how PRIMUS's seven science fields help to assess them. 
For high-redshift galaxy samples, we obtained statistically significant differences (up to 3-$\sigma$) between the clustering signal within COSMOS and other fields, though the samples were selected differently in S14 than in this paper.  

In Table~\ref{tab:HODresults3}, we list inferred HOD parameters for the ``nocosmos" samples in M15, which are like the default ones but exclude the COSMOS field, which has multiple rare large structures.  
Compared to the results in Section~\ref{sec:results}, the inferred halo-model parameters, especially at high redshift, are slightly different here than in Table~\ref{tab:HODresults}. 
For example, the inferred high-redshift halo masses are approximately 20\% ($0.1~\mathrm{dex}$) lower when COSMOS is excluded.  The other model parameters are very similar.

\begin{table*}[h]
\caption{Halo Occupation Distribution Results Excluding COSMOS Field}
 \centering
 \begin{tabular}{ l | c c c c c c }
   \hline
   sample & $\mathrm{log}M_\mathrm{min}$ & $\langle \mathrm{log}M_h\rangle$ & $\sigma_{\mathrm{log}M}$ & $M_1/M_\mathrm{min}$ & $\alpha$ & $f_\mathrm{sat}$ \\
   \hline 
   M1 & 11.70$\pm$0.05 & 12.00$\pm$0.10 & 0.20$\pm$0.10 & 17$\pm$1 & 1.09$\pm$0.05 & 0.22$\pm$0.02 \\
   M2 & 12.10$\pm$0.10 & 12.38$\pm$0.10 & 0.20$\pm$0.10 & 17$\pm$1 & 1.11$\pm$0.05 & 0.16$\pm$0.01 \\
   M3 & 12.60$\pm$0.10 & 12.85$\pm$0.10 & 0.20$\pm$0.10 & 16$\pm$1 & 1.10$\pm$0.05 & 0.11$\pm$0.01 \\
   M4 & 12.25$\pm$0.10 & 12.53$\pm$0.10 & 0.20$\pm$0.10 & 13$\pm$1 & 1.10$\pm$0.05 & 0.13$\pm$0.01 \\
   M5 & 12.60$\pm$0.10 & 12.83$\pm$0.10 & 0.20$\pm$0.10 & 15$\pm$1 & 1.07$\pm$0.05 & 0.09$\pm$0.01 \\
   \hline
  \end{tabular}
  \begin{list}{}{}
    \setlength{\itemsep}{0pt}
    \item HOD results with mock galaxy catalogs for PRIMUS catalogs with the COSMOS field excluded. These results are analogous to those presented in Table~\ref{tab:HODresults}.
  \end{list}
 \label{tab:HODresults3}
\end{table*}

But what does it mean to exclude or include a region with seemingly anomalous structures (or voids)?  For a discussion of this issue, see Norberg et al.\ (2011) and Zehavi et al.\ (2011). 
One could attempt to assess the effects of COSMOS's structures while using mocks and numerical simulations that include a regions with various large structures, but that would require better knowledge of the mass and spatial extent of the structures in the field and is beyond the scope of this paper.

\end{document}